\documentclass[prc,twocolumn,floatfix,nofootinbib]{revtex4-1}

\usepackage{graphicx}% Include figure files
\usepackage{amssymb,amsmath,amstext,amsthm,amsfonts}
\usepackage{slashed}
\usepackage{epstopdf}

\newcommand{\onepion}{single-pion\ }

\newcommand{\GeV}{\; \mathrm{GeV}}
\newcommand{\MeV}{\; \mathrm{MeV}}

\newcommand{\dd}{\mathrm{d}}

\begin{document}

\title{Pion production in the MiniBooNE experiment}
\author{O. Lalakulich%
\email[Contact e-mail: Olga.Lalakulich@theo.physik.uni-giessen.de]}
\author{U. Mosel}
\affiliation{Institut f\"ur Theoretische Physik, Universit\"at Giessen, D-35392 Giessen, Germany}

\begin{abstract}
\begin{description}
\item[Background] Charged current pion production gives information on the axial formfactors of nucleon resonances. It also introduces a noticeable background to quasi-elastic measurements on nuclear targets.
\item[Purpose] Understand pion production in neutrino interactions with nucleons and the reaction mechanism in nuclei.
\item[Method] The Giessen Boltzmann--Uehling--Uhlenbeck (GiBUU) model is used for an investigation of neutrino-nucleus reactions.
\item[Results] Theoretical results for integrated and differential cross sections for the MiniBooNE neutrino flux are compared to the data. Two sets of pion production data on elementary targets are used to obtain limits for the neutrino-nucleus reactions.
\item[Conclusions] The MiniBooNE pion production data are approximately consistent with the Brookhaven National Laboratory elementary data if a small flux renormalization is performed while the Argonne National Laboratory input data lead to significantly too low cross sections. A final determination of in-medium effects requires new data on elementary (\emph{p,D}) targets.
\end{description}
\end{abstract}

\maketitle

\section{Introduction}

In 2009 an unexpected experimental result appeared in studying quasielastic neutrino scattering from nuclei.
The cross sections reported by the MiniBooNE collaboration~\cite{Katori:2009du,AguilarArevalo:2010zc} at $E_\nu<2\GeV$ were about $30\%$ higher than the cross section measured by old bubble-chamber experiments in the 1970s-1980s. This excess was described in terms of a value for the axial mass that was significantly higher than the world-average value of about 1 GeV. At first sight, the new experiments have the advantage of huge statistics
with millions of events recorded. One complication, however, arises from the fact that these experiments all use nuclei as targets. All the measurements are thus influenced by nuclear effects. Indeed, the surprising result for the axial mass has since then found an explanation: The cross section labeled by the experiment as 'quasielastic' (QE) contains in reality a significant admixture of 2p-2h excitations that account for the difference between the data and theory using the standard value ($\approx 1 \GeV$) for the axial mass \cite{Martini:2011wp,Nieves:2011yp}.

The quasielastic data themselves have not directly been measured but have been deduced from so-called quasielastic-like data by subtracting a background of 'stuck-pion' events, i.e.,\ events in which pions are first produced, but then reabsorbed again. This background was determined from calculations with an event generator. Thus, the final QE + 2p-2h data invariably contain some model dependence.

A broad comparison of generator predictions for QE and \onepion cross
sections has shown major discrepancies ($\approx 30$\% for QE, up to 100\% for pion production) between the predictions of various generators presently being used by neutrino experiments~\cite{Boyd:2009zz}. That is why the new, quite complete data  on charged~\cite{AguilarArevalo:2010bm} and neutral~\cite{AguilarArevalo:2010xt} pion production in charged current (CC) neutrino scattering pion production cross sections obtained by the MiniBooNE collaboration are so important and could be crucial in verifying our theoretical understanding not only of this process, but also of QE scattering\footnote{MiniBooNE tuned their generator for the calculation of the stuck-pion background to their data on charged and neutral pion production
in charged current (CC) neutrino scattering.}.

There were early theoretical attempts to describe neutrino-induced pion production on nuclei before the data were available \cite{Leitner:2006ww,Ahmad:2006cy,SajjadAthar:2008hi}. Later studies of the measured $1\pi$/QE ratios all suffer from the fact that the QE cross section in the denominator of this ratio did not contain the 2p-2h contributions present in the data; a notable exception of this is the work by Martini \emph{et al.}\ \cite{Martini:2009uj} who were the first to realize the importance of the 2p-2h excitations. The few detailed comparisons of theory with the new results of the MiniBooNE experiment seem to indicate that a similar problem as for QE scattering exists also for neutrino-induced pion production \cite{Leitner:2009ec,Lalakulich:2011ne}; the data were found to be considerably above the calculated cross sections. This raises the question of the compatibility of the new neutrino-nucleus data with the old neutrino-nucleon data and of possible in-medium effects on neutrino-induced pion production.

On the theoretical side, all (with the exception of Ref.\ \cite{Martini:2009uj}) of the presently available detailed calculations of pion production rely on the so-called impulse approximation
in which the neutrino interacts with only one target-nucleon at a time. The interactions in medium, apart from Paul-blocking and Fermi-motion, are
assumed to be the same as in free space. In such a framework it is easy to understand
that, e.g., the total absorption cross sections get slightly reduced in medium in
comparison with the elementary reaction
(i.e.,\ the cross section summed up over all nucleons in the target as if they were free)
simply as a consequence of Pauli blocking. Considerably larger effects appear for
semi-exclusive final states  which are strongly affected by
final state interactions (FSI), which can alter the signature of the event.
Thus, the correct simulation of both QE and pion production events requires a
model that is able to describe the elementary reactions as well as the final state interactions.

It is the aim of this paper to investigate, first of all, the compatibility of the old pion production data on protons or deuterium with the new data obtained on nuclear targets. Second, by detailed comparisons with data for \onepion production on a nuclear target we study how far a realistic impulse approximation model of these processes can go in explaining the pion production data. Any remaining discrepancies could then possibly be attributed to primary many-particle interactions.

In Sec.\  \ref{gibuu}  we give a sketch of the GiBUU model.
In Sec.\  \ref{miniboone} we compare our calculations, which use two different parametrizations for elementary pion production as lower and upper bounds,
with the MiniBooNE data.
In particular, we investigate the influence of FSI on the absolute values and
the shape of the  observed distributions. Possible origins of remaining discrepancies will then be discussed in Sec.\  \ref{sect:discreps}.
Our findings are summarized in Sec.\  \ref{conclusion}.
Details on elementary pion-production inputs are given in Appendix~\ref{app:pion-theory}.
The influence of the medium modification of baryon properties on the various cross sections
is discussed in Appendix~\ref{app:Oset}.

\section{GiBUU transport model \label{gibuu}}
MiniBooNE has reported the cross section for the so-called CC ``observed \onepion production'',
that is for events with 1 pion of a given charge (there are data for $\pi^+$ and  $\pi^0$)
and no other pions in the final state, independent of the initial neutrino interactions vertex.
Thus, to describe those events theoretically, one needs a model with all relevant channels included:
QE scattering, $\Delta$ and higher resonance production, background 1-pion production, and DIS.

All these reaction mechanisms are contained in GiBUU.
In this paper we, therefore, employ this model for the analysis of the data. It has been developed
as a transport model for nucleon-, nucleus-,  pion-, and electron induced collisions from
some MeV up to tens of GeV. Several years ago neutrino-induced interactions were
also implemented for the energies up to few GeV. Thus, we can study all kind of elementary collisions
on all kind of nuclei within the unified framework. The code is written in modular Fortran
and is available for download as open source \cite{gibuu}.
Recently the GiBUU code was extended to describe also the 2p-2h excitations \cite{Lalakulich:2012ac} and the DIS reactions \cite{Lalakulich:2012gm} for neutrino-induced reactions. Below we shortly describe the nuclear model implemented in the GiBUU code and used in this analysis. For more details on the GiBUU model see the review in Ref.~\cite{Buss:2011mx}.

\subsection{Elementary neutrino interactions \label{elementary}}

\begin{figure*}[!htb]
\begin{minipage}[c]{0.48\textwidth}
\includegraphics[width=\textwidth]{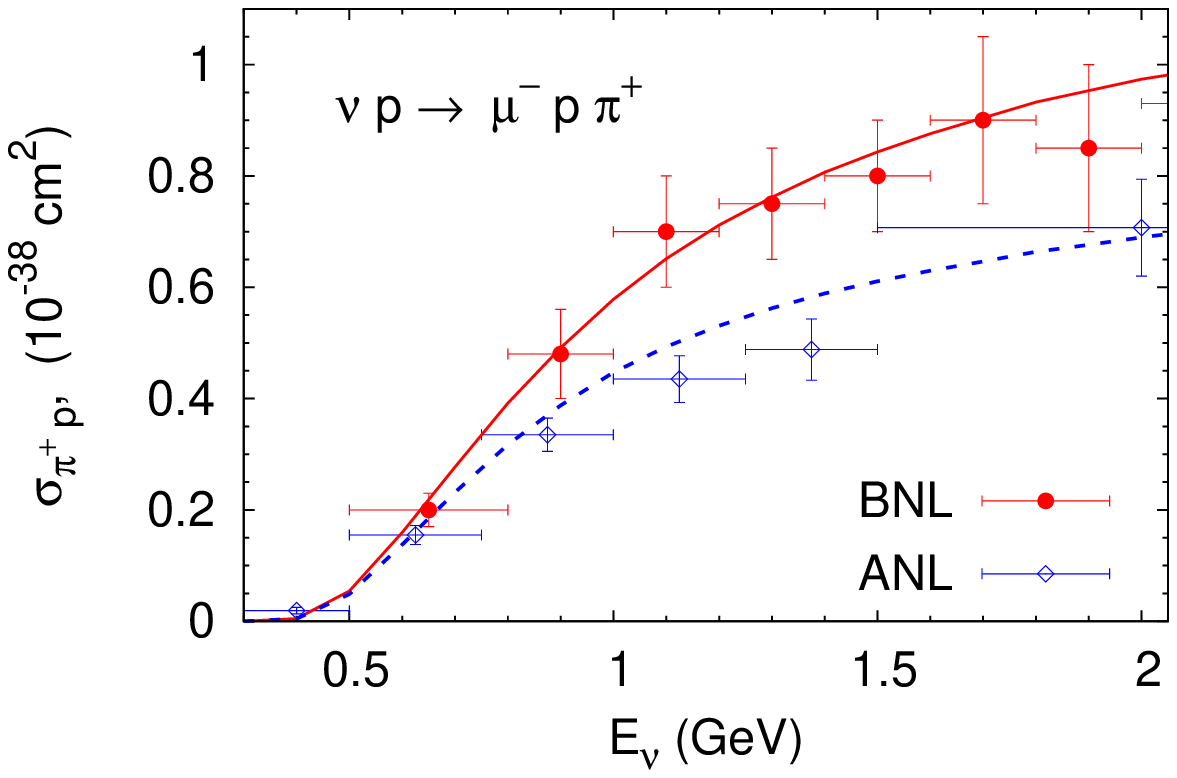}
\end{minipage}
\hfill
\begin{minipage}[c]{0.48\textwidth}
\includegraphics[width=\textwidth]{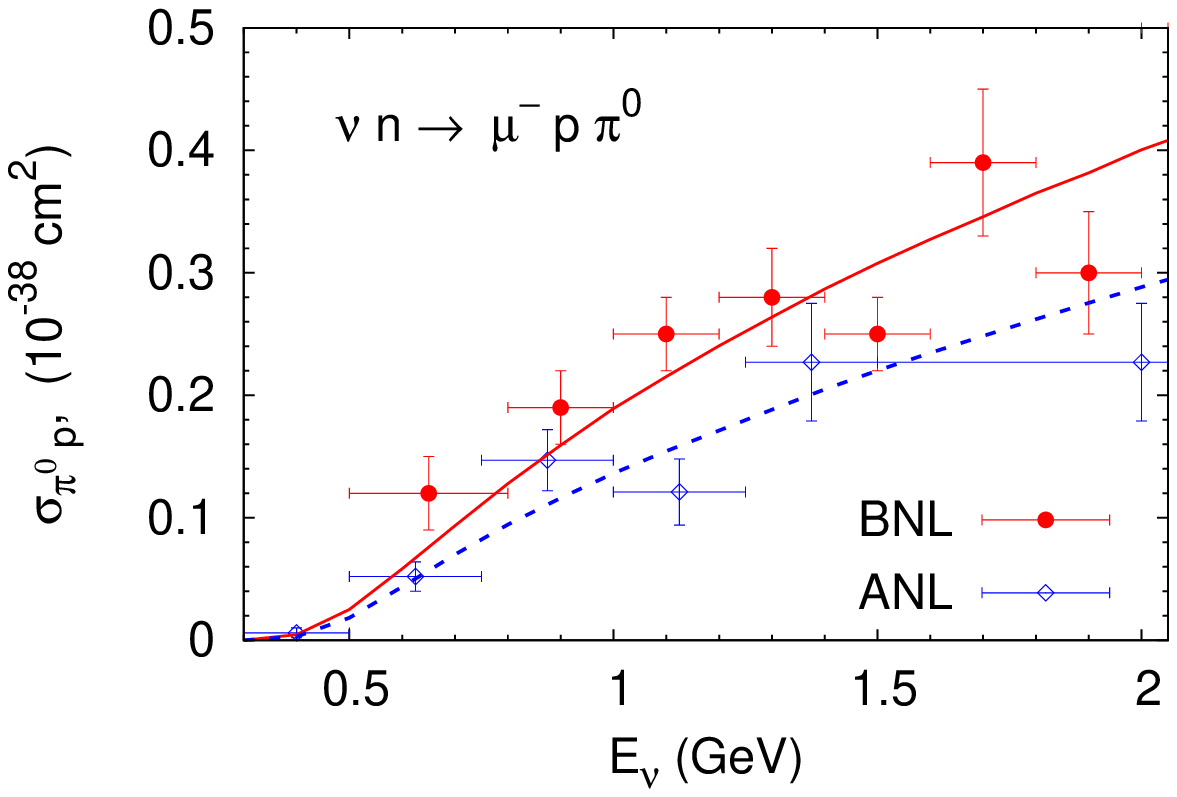}
\end{minipage}
\caption{(Color online) Single-pion production cross section on proton and neutron targets obtained in the BNL \cite{Kitagaki:1986ct} (circles; solid curve) and the ANL experiments \cite{Radecky:1981fn}  (diamonds; dashed curve). The curves give the lower (ANL-tuned) and upper (BNL-tuned) boundaries on the elementary input as used in GiBUU.}
\label{fig:ANLBNL1pidata}
\end{figure*}
A necessary input into all transport codes are the elementary reaction cross sections.
Neutrino and antineutrino interactions with nucleons may result in several different channels.
At hadronic invariant masses below the \onepion production threshold, $W<1.08 \GeV$, only the QE processes
$\nu n \to \mu^- p$ and $\bar\nu p \to \mu^+ n$ are possible. They are described within the standard
form factor formalism as outlined in our previous publications \cite{Leitner:2006ww,Leitner:2008ue}.
We emphasize, that for the axial mass we use the world average value
$M_A=0.999 \GeV$~\cite{Kuzmin:2007kr} in all calculations.

At $W>1.08 \GeV$ the \onepion production channel opens. The biggest contribution comes from
the $\Delta$ [$P_{33}(1232)$] resonance, at least up to invariant
pion-nucleon masses of $W \approx 1.5 \GeV$. In Refs.\ \cite{Leitner:2006ww,Leitner:2008ue} we had used the ANL data on the proton \cite{Barish:1978pj,Radecky:1981fn} to pin down the details of the axial coupling to the $\Delta$ resonance.  We had used the ANL data as our standard input because here also the absolute values of the cross section for $\dd\sigma/\dd Q^2$ were given.

In this paper we now also show results obtained by using the BNL pion production cross sections \cite{Kitagaki:1986ct} as input in order to explore the consistency of both input data sets with the recent MiniBooNE data. Fig.\ \ref{fig:ANLBNL1pidata} shows that the data obtained in the BNL experiment are generally higher than those from the ANL experiment. It has been argued in \cite{Graczyk:2009qm} that the two data sets are statistically consistent when taking into account their individual errors and possible errors in the flux normalization. A joint analysis of both data sets was also performed in \cite{Hernandez:2010bx}. In order to obtain an estimate for the systematic uncertainty in the calculations we fit here both of these data sets separately. The curves in Fig.\ \ref{fig:ANLBNL1pidata}a show the theoretical description as obtained from such a fit and as used as input in GiBUU. In Appendix \ref{app:pion-theory} we give the relevant parameters.

With increasing $W$, excitations of resonances with higher masses, followed by their subsequent decays,
give increasing contributions. For the prominent $\Delta$ resonance, the $\Delta\to N \pi$ channel
saturates the one-pion width at the $99\%$ level.  The higher resonances, however, can also produce two and more pions as well as other mesons.  In the GiBUU code 13 of 19 resonances having three- and four-star ratings in the Review of Particle Properties by the Particle Data Group  are accounted for\cite{Beringer:1900zz}. The remaining six are unaccounted for because their electromagnetic properties as unknown. The resonance production is described phenomenologically, as presented in \cite{Leitner:2006ww,Leitner:2008ue}.
Namely, for the higher lying resonances the vector couplings are taken directly over from the MAID analysis \cite{MAID} so that the model is consistent with all results for electromagnetic interactions. The axial couplings are obtained from PCAC.

In the same $W$ region,  non-resonant background gives a noticeable  contribution.
These processes are discussed in \cite{Leitner:2008ue}. The default parameters of the model are tuned in such
a way as to describe the ANL data on $p\pi^0$ and $n\pi^+$ final states
in CC neutrino scattering on neutron.  In this paper we also show results obtained by using the
generally higher BNL pion production cross sections as input. Both ANL- and BNL-tuned inputs are shown in Fig.~\ref{fig:ANLBNL1pidata}
together with the corresponding data.

DIS events are included by using the \textsc{PYTHIA} event generator.
The produced particles are then further propagated by the FSI treatment of GiBUU. While the DIS
picture is applicable at high or at least moderate squared momentum transfer $Q^2$,
at low $Q^2$ \textsc{PYTHIA} treats the interaction within the vector meson dominance
model as the production of vector mesons and their subsequent decays.

\subsection{Description of initial nucleus \label{init-nucleus}}

GiBUU describes the struck nucleus  as a collection of off-shell nucleons.
Each nucleon is bound in a mean-field potential, which on average describes the many-body interactions with the other
nucleons. This potential is parametrized as a sum of a Skyrme term depending only on density, and
a momentum--dependent contribution.
The phase space density of nucleons is treated within a local Thomas-Fermi approximation.
At each space point the nucleon momentum distribution is given by a Fermi sphere, whose radius
in the momentum space is determined by the local Fermi momentum, which depends on the nucleon
density.  Within this picture, contrary to the  Fermi gas model with constant Fermi momentum
(the global Fermi gas model), the nucleon position and momentum are correlated. This leads to
a smoother momentum distribution with somewhat more strength at lower momenta and to smoother
nucleon spectral functions.
A more detailed description, comparison with the global Fermi gas model
and numerical values of the parameters used are given in~\cite{Leitner:2008ue},
in particular Fig.\ 6 there.

All calculations reported in this paper are based on the impulse approximation in which the incoming lepton interacts with a single bound nucleon, with the
interaction vertex being the same as in the case of a free nucleon. The vertices for the
quasielastic scattering, resonance excitations and background contribution are given
in~\cite{Leitner:2006ww,Leitner:2008ue}, those for 2p-2h excitations in \cite{Lalakulich:2012ac} and for DIS in \cite{Lalakulich:2012gm}. In the publications quoted we have shown that such an approach can give a very good description of the measured inclusive double-differential (with respect to the outgoing muon) cross sections. Incorporated into GiBUU it can make also detailed predictions for knock-out particle spectra and multiplicities.

\subsection{Pion production}
The \onepion production cross section is then given by (cf.\ \cite{Buss:2011mx}, sec.\ 4.2.1)
\begin{widetext}
\begin{equation}   \label{piprod}
\dd \sigma^{\nu A \to \ell'X\pi} = \int \dd E \int \frac{\dd^3p}{(2\pi)^3} P(\mathbf{p},E) f_{\rm corr}\, \dd\sigma^{\rm med} \, P_{\rm PB} (\mathbf{r},\mathbf{p}) F_\pi(\mathbf{q}_\pi, \mathbf{r}) ~.
\end{equation}
Here $f_{\rm corr}$ is a flux correction factor $f_{\rm corr} = (k \cdot p)/(k^0p^0)$; $k$ and $p$ denote the four-momenta of the neutrino and nucleon momentum, respectively. The factor $P_{\rm PB}$ describes the Pauli-blocking and $P$ describes the hole spectral function, in the local Thomas-Fermi model given by
\begin{equation}
P(\mathbf{p},E) = g \int\limits_{\rm nucleus} \!\!\!\dd^3r \,\Theta\left[p_\mathrm{F}(\mathbf{r}) - |\mathbf{p}|\right] \Theta(E) \delta\left(E - m^* + \sqrt{\mathbf{p}^2 + {m^*}^2}\right)~,
\end{equation}
here $p_\mathrm{F}(\mathbf{r})$ is the local Fermi momentum  given by the local Thomas Fermi model. In this spectral function all effects of the nucleon potential are assumed to be contained in the effective mass $m^*$ \cite{Buss:2011mx} which depends on location and momentum of the nucleon. Finally, the factor $F_\pi(\mathbf{q}_\pi, \mathbf{r})$ in (\ref{piprod}) describes the effects of all the FSI contained in GiBUU, as briefly described in sec.\ \ref{fsi}.

The cross section $\dd\sigma^{\rm med}$ in Eq.\ (\ref{piprod}) describes the elementary pion production cross section. It can be separated into terms depending on the number of nucleons involved in the initial interaction:
\begin{equation}    \label{IA}
\dd\sigma^{\rm med} = \dd\sigma_{1p1h1\,\pi}^{\rm med} + \dd\sigma_{2p2h\,1\pi}^{\rm med} + ... ~.
\end{equation}
\end{widetext}
Here the first term contains the standard cross section for $1\pi$ production on one nucleon as described in detail in \cite{Leitner:2006sp,Leitner:2006ww,Leitner:2008ue} as briefly summarized in Appendix \ref{app:pion-theory} together with the parameters used. It is proportional to nuclear density $\propto \rho$. The second term contains production processes that involve 2 or more nucleons; it is at least $\propto \rho^2$.

The superscript 'med' in Eq.\ (\ref{IA}) is meant to indicate that the spectral functions of the particles involved can change in medium due to binding.
In all calculations we use the impulse approximation, i.e.,\ only the first term in Eq.\ (\ref{IA}), with in-medium changes for the width of the $\Delta$ resonance taken from the work of Oset and Salcedo (OS) \cite{Oset:1987re}. The separation into 1p-1h and 2p-2h contributions then involves a subtlety: if we use the OS result  for the $\Delta$ width in the first term, in a diagrammatical approach this already involves terms of 2p-2h nature. In sec. \ref{sect:discreps} we will discuss the possible influence of higher-order (in density) processes.

All our calculations are presented for a $\mathrm{CH}_2$ target.
Charged  pions can be initially produced on the proton as well as on the carbon nucleus.
In the latter case they undergo FSI. Neutral pions are initially  produced on carbon, except for
a negligible contribution from DIS scattering on proton at higher energies.

\subsubsection{In-medium modifications of nucleons and resonances \label{modif-default}}

As outlined in \cite{Leitner:2008ue}, each outgoing baryon is described by its spectral function.
The latter depends on its self-energy, which in turn depends on four-momentum and
position of the particle in the nucleus. The spectral function of a free resonance is
a Breit-Wigner distribution with a width equal to the free width of the corresponding baryon resonance.
In lepton-nucleus interactions the outgoing particles are initially produced inside the nucleus,
and, thus, are off-shell and bound in the mean-field potential described above.
In medium, on one hand, the width is lowered due to Pauli blocking.
Indeed, consider  e.g.\ a nucleon resonance, decaying into a pion nucleon pair. This decay is
Pauli-blocked if the nucleon momentum is below the Fermi momentum. On the other hand, the width
is increased by the collisions inside the nucleus. For example, via the processes
$\Delta N \to NN$ and $\Delta NN \to NNN$ the $\Delta$ can disappear without producing a pion.
Secondary pion production is also possible, namely via the process $\Delta N \to \pi NN$.
These processes, as well as elastic scattering $\Delta N \to \Delta N$, all contribute to the
collisional width. It is essential here to treat the collisional broadening consistently with the corresponding collision terms. This means, for example, that whenever contributions from e.g.\ $\Delta NN \to NNN$ to the $\Delta$ width in the OS model are taken into account, then also the corresponding three-body collisions have to be taken into account. This is, indeed, the case in GiBUU.

A theory to calculate this medium modification was worked out by Oset and Salcedo~\cite{Oset:1987re}.
These authors evaluated the $\Delta$ self-energy using a many-body expansion in terms of particle-hole and $\Delta$-hole excitations. Quasi--elastic corrections, two- and three-body absorption are included in this width.  In Appendix \ref{app:Oset} we discuss the influence of the OS broadening of the $\Delta$ resonance on the pion yields in some detail.

The GiBUU code can be run with all the prescriptions for medium modifications mentioned above.
The default option in GiBUU is using the OS result for the $\Delta$ width and neglecting medium modifications for higher resonances.
The latter give  only a relatively minor contribution and using the full collisional width is computationally expensive.
We will discuss the influence of the various treatments of spectral functions on the final, observable results in some more detail in Sec.~\ref{sect:discreps} and Appendix~\ref{app:Oset}.

\subsection{Final State Interactions \label{fsi}}

In experiments with nuclear target the outgoing hadrons are masked by the final state
interaction: protons can rescatter in a nucleus and produce additional mesons or knock-out another nucleons.
Pions, that were originally produced through a weak excitation of $\Delta$ resonance, can be absorbed in the nucleus
or converted to other mesons. Thus, the inevitable presence of the FSI washes out the true origin of the event and makes an
experimentally observed signal different from what one would expect for the  scattering on a free nucleon. FSI can decrease the cross sections as well as significantly modify
the shapes of the final particle spectra. In the present study FSI will,
for example,  change the signature of initial pion production; through FSI this process can lead to knock-out nucleons which mimic QE scattering.

The restoration of the true cross section of the process of interest, therefore,
cannot be achieved by pure experimental means and crucially relies on theoretical modeling
implemented in the event generators. For a reliable interpretation of experiments one thus
needs a model which provides a realistic description of both initial neutrino-nucleus interaction
and the final state interaction of the produced hadrons.

In GiBUU FSI are implemented by solving the semi-classical Boltzmann-Uehling-Uhlenbeck (BUU) equation.
It describes the dynamical evolution of the phase space density for each particle species
under the influence of the mean field potential, introduced in the description of
the initial nucleus state. Equations for various particle species are coupled through this mean field and
also through the collision term. This term explicitly accounts for changes in
the phase space density caused by elastic and inelastic collisions between particles.
For a more detailed discussion of FSI see Ref.~\cite{Buss:2011mx}.

\subsection{Model validation}
The GiBUU model has extensively been tested and has been found to give a good description of experimental data for very different reaction mechanisms \cite{Leitner:2009ke}. For the present investigation comparisons with pion-nucleus and with photon-nucleon reactions are most relevant. The formalism used for their description is the same as that just described in the preceding section. Only the neutrino as incoming particle has to be replaced by either a pion or a photon. For both reaction types, however, the FSI are the same.

Tests of pion absorption on nuclei have been passed by GiBUU~\cite{Buss:2006vh}. However, these total absorption cross sections are not very sensitive to details of the pion's interactions with the surrounding nucleus. More sensitive to the details of the pion interactions with the nuclear medium, and particularly relevant for the neutrino-induced $\pi^0$ production on nuclei, is the
double-charge exchange reaction of pions with different nuclear targets. GiBUU calculations achieve a good quantitative
agreement with the extensive data set obtained at LAMPF for incident pion kinetic energies of $120-180\MeV$ ~\cite{Buss:2006yk}.

\begin{figure*}[!hbt]
\includegraphics[width=0.7\textwidth]{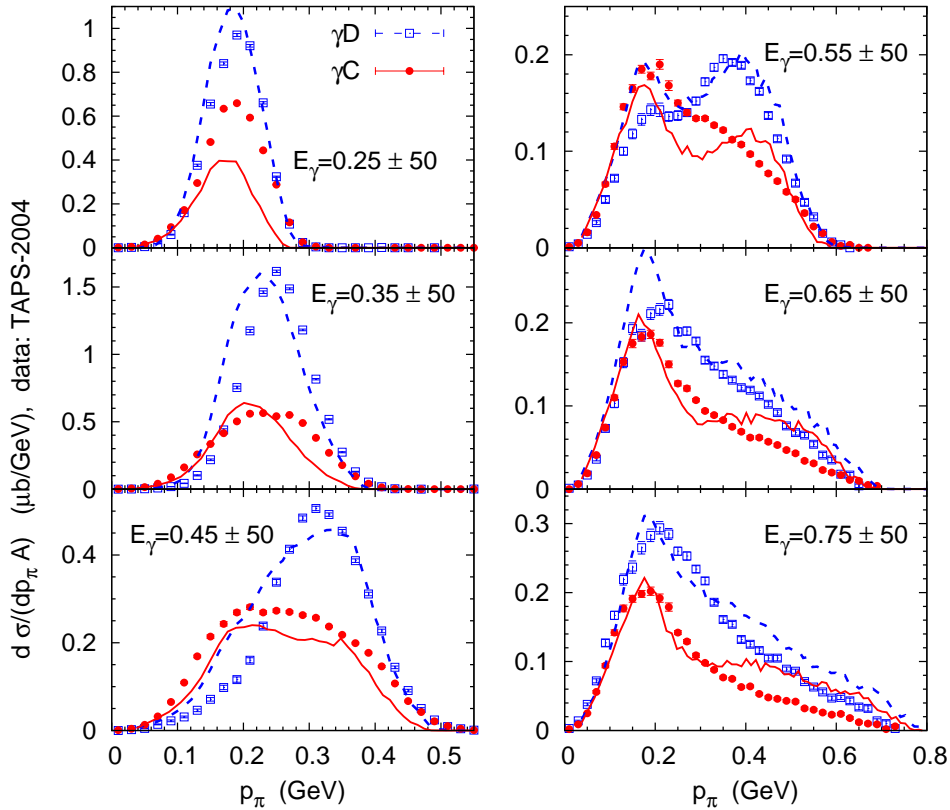}
\caption{(Color online) Momentum distributions of the outgoing pions for inclusive $\pi^0$ production in scattering
 of photons of energies $0.2-0.8\GeV$ off D and C nuclei. The dashed curve shows the calculations for a D target, the solid one that for C. The data are from \cite{Krusche:2004uw}. They contain coherent pion production, which is not included in the calculations; it plays a role only on the low-momentum side of the peak. }
\label{fig:Lehr-pi0}
\end{figure*}

Closely related to neutrino-induced reactions is the photoproduction of pions.
On the nucleon (i.e.,\ without final state interactions) these reactions  directly test the vector part of
the pion-production vertex.  On nuclei this reaction tests the nuclear dynamics of pion propagation throughout the nucleus.
GiBUU describes the dataset for photoproduction of neutral pions~\cite{Krusche:2004uw}
on nuclei for photon energies up to $0.8\GeV$ quite well \cite{LehrDiss:2003}.

As an illustration, that will become relevant for the later discussions, Fig.~\ref{fig:Lehr-pi0} shows measured and calculated pion momentum distributions for $\pi^0$ photoproduction off D and C nuclei. The shapes of the experimental distributions change significantly when going from deuterium (which is nearly equivalent to production before FSI) to  C (corresponding to production after FSI). The main effect is a strong absorption of pions around momenta of 0.3 GeV due to the excitation of the $\Delta$ resonance and its subsequent pionless decay. GiBUU calculations\footnote{These were originally done in \cite{LehrDiss:2003} and are now checked with the present version of the code} reproduce this behavior and show a generally good agreement with the  data (solid and dashed lines in Fig.\ \ref{fig:Lehr-pi0}); in the calculations the absorption around 0.3 GeV is indeed due to $N\Delta \to NN$ or $NN\Delta \to NNN$ collisions. The remaining discrepancies between theory and experiment for C give an indication of the systematic errors in the GiBUU calculations. We also note here that the calculations shown in Fig.~\ref{fig:Lehr-pi0} all use the collision-broadened width as given by Oset and Salcedo \cite{Oset:1987re}. Using the free width instead would yield significantly too large cross sections at the peak position.

Electroproduction data for pions are very sparse, but the HERMES data at 28 GeV and the recent JLab data at 4 - 5.8 GeV are again reproduced fairly well \cite{Gallmeister:2007an,Kaskulov:2008ej}. Also the total inelastic electron scattering data are described very well as shown in Ref.\ \cite{Leitner:2008ue}. There, in Figs.\ 9 and 10, it is also shown, that the inclusion of the medium-modified width
gives a better description of the double differential electron scattering cross section. In particular,
the QE peak is lowered and the cross section in the dip region between the QE and Delta peaks is enhanced in agreement with the data.

Therefore, all the neutrino-induced calculations in this paper are performed using the expressions for the in-medium width of the $\Delta$ resonance as given in \cite{Oset:1987re}. The general effect of this in-medium correction is again a significant lowering of the pion production cross section at its $\Delta$ peak and a broadening of the latter. These changes are thus the same as seen in the description of the photonuclear data. They depend only on the in-medium properties of the $\Delta$ resonance and not on its population mode and should thus be taken into account also in calculations of neutrino-induced pion production. For neutrinos these effects are discussed in more detail in Appendix~\ref{app:Oset}.

\section{MiniBooNE CC charged and neutral pion production}
\label{miniboone}

In this section we compare our model predictions with the MiniBooNE  data on pion production.

The MiniBooNE energy flux \cite{AguilarArevalo:2008yp} peaks at $0.6\GeV$ and becomes very small above $1.7\GeV$. We use this flux in all our calculations.

In the following discussions the data from~\cite{AguilarArevalo:2010bm,AguilarArevalo:2010xt} are plotted versus reconstructed neutrino energy, while the theoretical curves presented here are versus real energies.  Here we neglect this difference.\footnote{We have found that the energy reconstruction method used in the experiment \cite{AguilarArevalo:2010xt}, that relies only on the kinematics of the outgoing muon and pion, is quite quite reliable} For the later comparisons it is also essential to note that all the experimental cross sections for $\pi^+$ production were obtained with the full MiniBooNE flux. Thus, for positively charged pions the distributions with respect to muon or pion energy do not depend on any energy reconstruction scheme. This is not so for the $\pi^0$ data where a cut for the (reconstructed) neutrino energies was imposed: only neutrino energies between 0.5 and 2.0 GeV were taken into account \cite{AguilarArevalo:2010xt}. Since this cut is not possible without a generator-based reconstruction procedure the $\pi^0$ data may thus contain some model dependence. We also note that all our earlier calculations published in Refs.\ \cite{Leitner:2008wx,Leitner:2009de,Leitner:2009ec,Lalakulich:2011ne} used the full MiniBooNE flux, without this cut, for all charge states. The net result of using now the neutrino energy-window for $\pi^0$ is an increase of the calculated flux-averaged cross sections for neutral pions compared to the earlier calculations.

\subsection{Energy-dependence of cross sections}
\label{miniboone-integrated}
\begin{figure*}[!hbt]
\includegraphics[width=\textwidth]{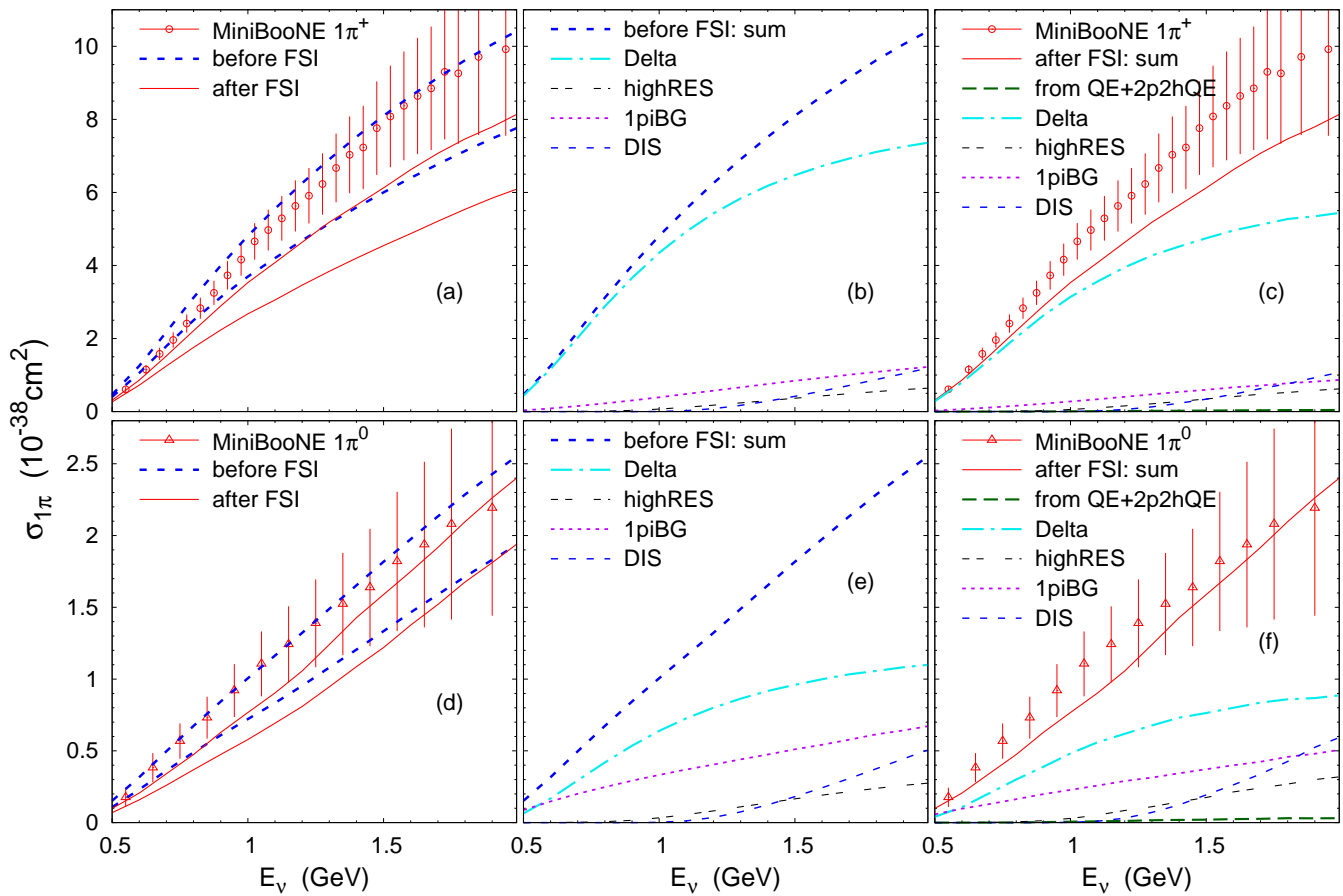}
\caption{(Color online) Integrated cross section for  1$\pi^+$ (a,b,c) and
1$\pi^0$(d,e,f) CC production versus neutrino energy. (a) and (d) contain the results of calculations using ANL- (lower curves) and the BNL-tuned (upper curves) elementary input both before and after FSI.
(b) and (e) show the contributions of various event origins to the calculated cross sections before FSI.
(b) and (f) show the same after FSI. Here the BNL-tuned input has been used. Data are from~\cite{AguilarArevalo:2010bm,AguilarArevalo:2010xt}.}
\label{fig:MB-lepton-Enu-QEDelta}
\end{figure*}

We start our discussion with a comparison and discussion of energy-unfolded cross sections for pion production.
The cross sections for \onepion production versus neutrino energy for CC $1\pi^+$ and $1\pi^0$ production
are  shown in Fig.~\ref{fig:MB-lepton-Enu-QEDelta}.

\subsubsection{$\pi^+$ production.} Figures~\ref{fig:MB-lepton-Enu-QEDelta}a,b,c show the results for $1\pi^+$ production.
Immediately noticeable in Fig.\ \ref{fig:MB-lepton-Enu-QEDelta}a is that the BNL elementary input leads to a significantly larger cross section than the
ANL input also for a nuclear target (CH$_2$ in this case), simply reflecting that fact that already the elementary BNL cross sections are about 30\% larger than those from the ANL experiment (see Fig.\ref{fig:ANLBNL1pidata}). In the following we consider the cross sections obtained with the BNL and ANL input as upper and lower boundaries, respectively, that illustrate the systematic uncertainties in the final theoretical results.
For both inputs the calculated cross section after FSI (solid lines) is about 30\% lower than the
corresponding cross section before FSI (dashed lines).
For the cross section after FSI, which is the one to be compared to the data, the slope of the upper boundary (BNL input) is about the same as the
slope of the data (within experimental errors).  The lower boundary (ANL input), on the other hand, is flatter than the data. As a consequence its difference to the data increases with energy.

Figure~\ref{fig:MB-lepton-Enu-QEDelta}b shows the origin of the $1\pi^+$ events before FSI. Most of them
come from initial  $\Delta$ resonance production and its following decay (dash-dotted line). This channel is dominant up to about 0.9 GeV incoming neutrino energy. Some  events are background ones (dotted line). There is also a small contribution from higher resonances, and at $E_\nu> 1.2 \GeV$ DIS processes start to play an increasing role.

Figure~\ref{fig:MB-lepton-Enu-QEDelta}c shows the origin of the $1\pi^+$ events after FSI.
As can be seen by comparing panels (b) and (c), FSI noticeably decrease the $\Delta$-originated  \onepion production due to the absorption $N\Delta\to NN$; the similar process is possible also for other resonances. Once a pion is produced, independent of its origin, it may also undergo a charge-exchange $\pi^+ n \to \pi^0 p$ process, which depletes the $\pi^+$
channel as the dominant one. The ensuing reduction of the $1\pi$-background channel due to FSI is also noticeable.
Other possibilities for pions to disappear include $\pi N \to \omega N$, $\phi N$, $\Sigma K$, $\Lambda K$.

A minor amount of pions comes from the initial QE vertex (long-dashed line),
which is only possible due to FSI, when the outgoing proton is rescattered.
Here the main contribution is from  the $p N \to N' \Delta \to N' N^{''} \pi$ reaction.
Other possibilities to create pion during the FSI would be $\omega N \to \pi N$, $\phi N \to \pi N$,
$\pi N \to \pi \pi N$.

When comparing to experiment, one sees that the lower boundary (ANL input) lies considerably ($\approx 40-60\%$) below the data  with the discrepancy increasing with neutrino energy. For the upper boundary (BNL input), on the other hand,  the discrepancy to the data becomes significantly smaller. However, even this theoretical cross sections is still by about 15-20\% below the data.

\subsubsection{$\pi^0$ production.} Figure~\ref{fig:MB-lepton-Enu-QEDelta}d shows the results for $1\pi^0$ production.
Here the curves before and after FSI are not so different. This is mainly due to the fact, that
side feeding from $\pi^+ n \to \pi^0 p$, which decreases the charged pion output, simultaneously
increases the neutral pion output.
The reverse process gives only minor relative contributions, because the initial $\pi^0$ production cross section is around 5 times lower.
This is, however, partly compensated by charge exchange to the $\pi^-$ channel through
$\pi^0 n \to \pi^- p$ and other channels mentioned above for charged pion production.

Figures~\ref{fig:MB-lepton-Enu-QEDelta}e,f show the origin of the $1\pi^0$ events before and after FSI, respectively.
Here the $\Delta$ channel is less important than in the case of charged pions,
while background processes, and at higher energies higher resonances  and DIS play a relatively larger role than for $\pi^+$.
Comparing these calculations with experiment one sees that the lower boundary (ANL input)  lies again about 40\% below the data at 1 GeV. This discrepancy becomes smaller at higher energies due to a change in slope at about 1.2 GeV, which comes from the opening of higher resonance excitations and DIS events. The upper boundary (BNL input) also shows this bend, but is much closer to the data (within about 20\% at 1 GeV, and within the experimental uncertainties at higher energies).

\begin{figure}[!hbt]
\includegraphics[width=\columnwidth]{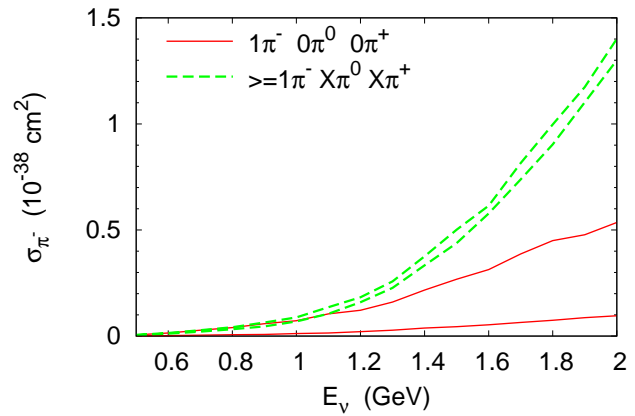}
\caption{(Color online) Integrated cross section for CC $\pi^-$ production
versus neutrino energy. The solid curves give the results for 1 $\pi^-$ and no other pions, the dashed curves those for any number of pions, including at least 1 $\pi^-$. In each case the upper curve gives the results obtained with the BNL input and the lower ones those with the ANL input cross sections.}
\label{fig:MB-lepton-Enu-QEDelta-piminus}
\end{figure}
\subsubsection{$\pi^-$ production.}
Our predictions for $\pi^-$ production are shown in Fig.~\ref{fig:MB-lepton-Enu-QEDelta-piminus}.
Without FSI the production of $\pi^-$ is negligible, a few events may only come from higher
mass resonances in the processes like $R^+ \to p \rho^0 \to p \pi^+ \pi^- $ or
$R^+ \to p \Delta^0 \to p p \pi^-$ or from DIS (the curves are not shown because they would be indistinguishable from zero).
Cross sections for both $1\pi^-$ (solid curve), which is defined as one $\pi^-$ and no other pions, depends significantly on the elementary input used; its lower to upper boundaries (ANL to BNL input) differ by a factor of 5.
The uncertainty band for multi-$\pi^-$ production, which is defined as events with at least one  $\pi^-$ and any number of pions of other charges, is significantly narrower. Events with more than one pion show up only at neutrino energies above $\approx 1$ GeV. The cross sections for $\pi^-$ production appear to be around 10 - 20 times smaller than those
for $1\pi^+$. However, with the current statistics of the MiniBooNE experiment they should be measurable.

\subsection{Lepton observables}

\begin{figure*}[!hbt]
\begin{minipage}[c]{0.48\textwidth}
\includegraphics[width=\textwidth]{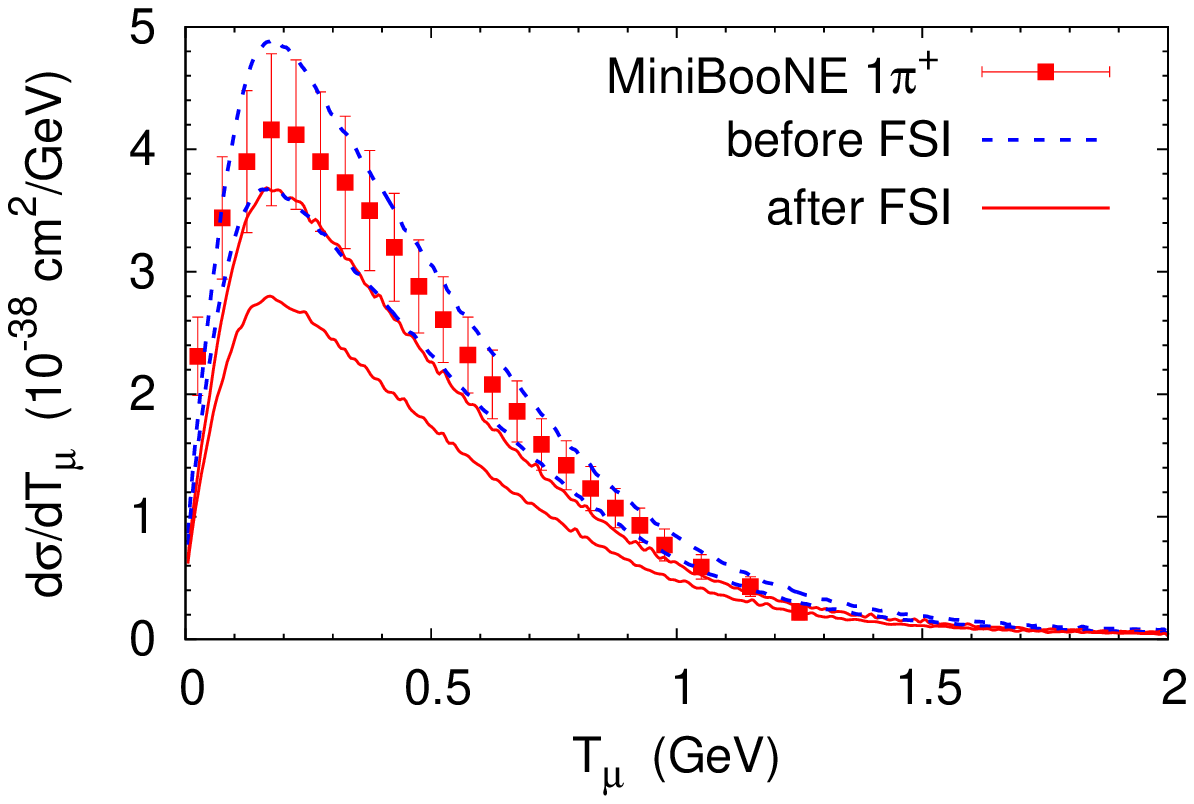}
\end{minipage}
\hfill
\begin{minipage}[c]{0.48\textwidth}
\includegraphics[,width=\textwidth]{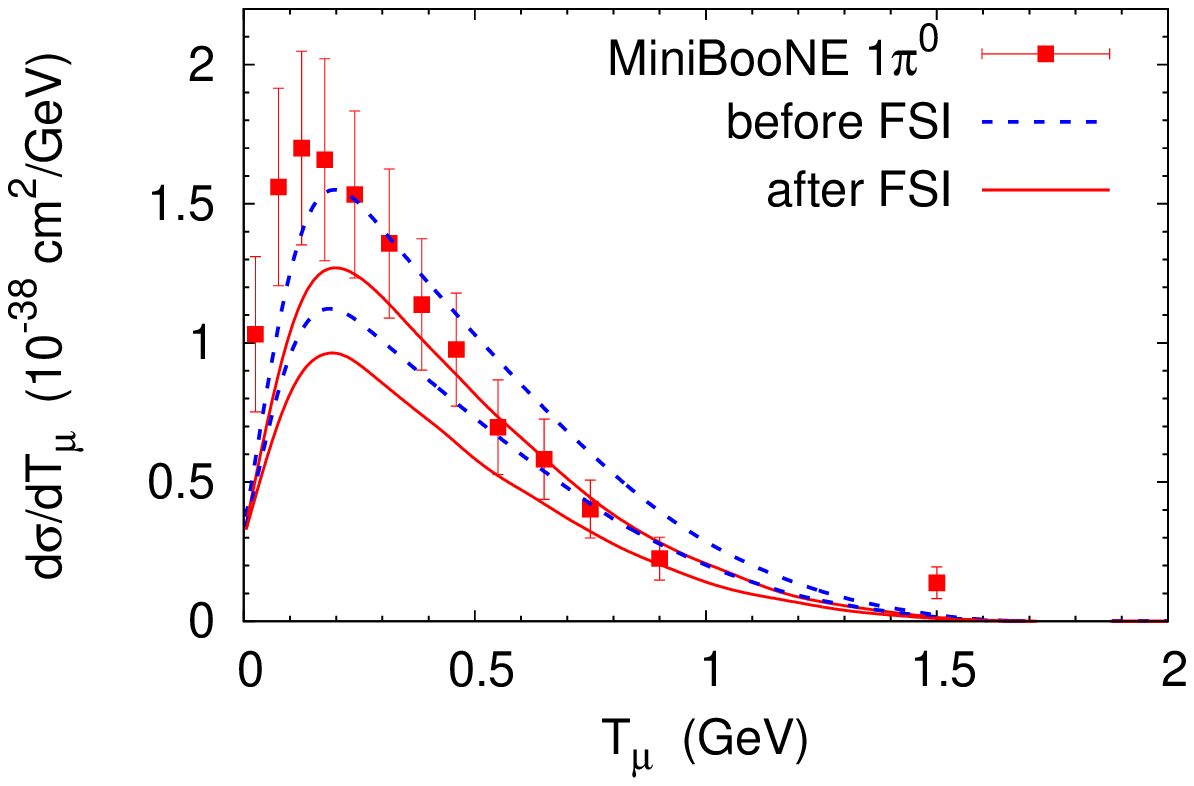}
\end{minipage}
\caption{(Color online) Kinetic energy distributions of the outgoing muons
for the  1$\pi^+$ and 1$\pi^0$ production in the MiniBooNE.
Data are from~\cite{AguilarArevalo:2010bm,AguilarArevalo:2010xt}. The dashed curves give the results before FSI, the solid curves those with all FSI included.}
\label{fig:MB-lepton-Ekin}
\end{figure*}

\begin{figure*}[!hbt]
\begin{minipage}[c]{0.48\textwidth}
\includegraphics[width=\textwidth]{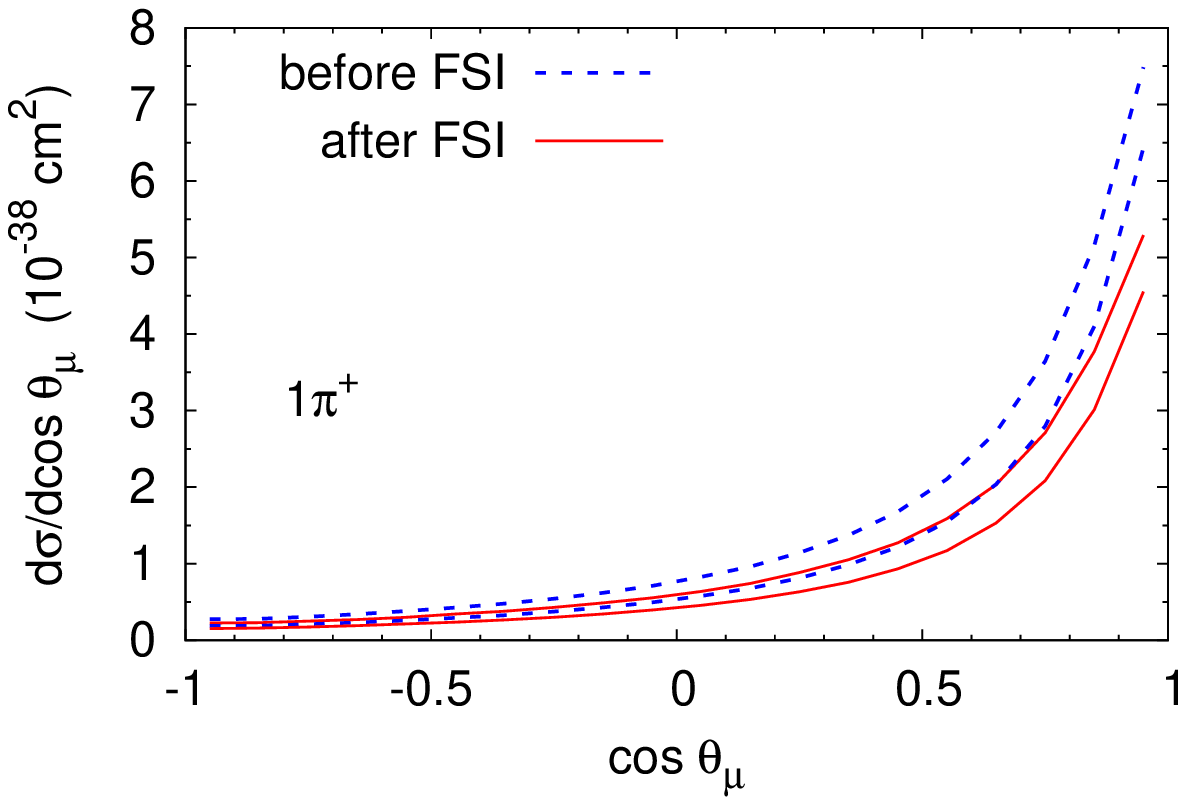}
\end{minipage}
\hfill
\begin{minipage}[c]{0.48\textwidth}
\includegraphics[width=\textwidth]{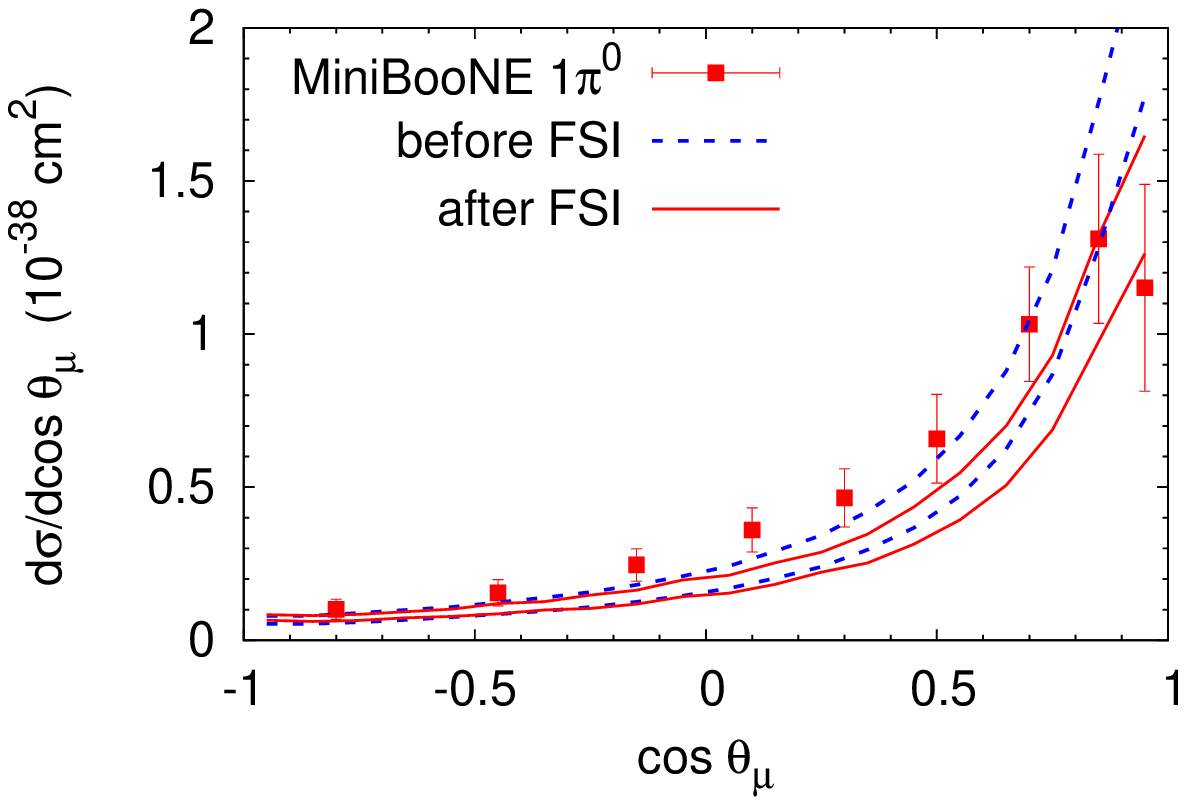}
\end{minipage}
\caption{(Color online) Angular distribution of the outgoing muons for the 1$\pi^+$ and 1$\pi^0$ production in the MiniBooNE.
Data are from~\cite{AguilarArevalo:2010xt}. The curves are as in Fig.\ \ref{fig:MB-lepton-Ekin}.}
\label{fig:MB-lepton-coss}
\end{figure*}

The MiniBooNE collaboration has recently also published distributions versus muon kinetic energy, $d\sigma/dT_{\mu}$, muon angle with respect
to the neutrino  beam direction, $d\sigma/d\cos\theta_{\mu}$, and squared four-momentum transfer, $d\sigma/dQ^2$~\cite{AguilarArevalo:2010bm}. While the former two distributions in principle do not depend on energy reconstruction the latter distribution does depend on it because $Q^2$ has to be reconstructed. There is, however, a subtle difference here in the way how the data for $\pi^+$ and $\pi^0$ production were obtained. Whereas the former indeed are averaged over the full neutrino flux as given in \cite{AguilarArevalo:2008yp} the data for $\pi^0$ were actually restricted to the incoming (reconstructed) neutrino energy range from 0.5 to 2.0 GeV and thus contain some model dependence. In our calculations we have taken this restricted energy range into account by normalizing the experimental flux distribution to 1 over only this limited energy range.\footnote{In our earlier calculations \cite{Lalakulich:2011ne} we had used the full MiniBooNE flux, without any restrictions. This led to even lower cross sections.}

\begin{figure*}[!hbt]
\begin{minipage}[c]{0.48\textwidth}
\includegraphics[,width=\textwidth]{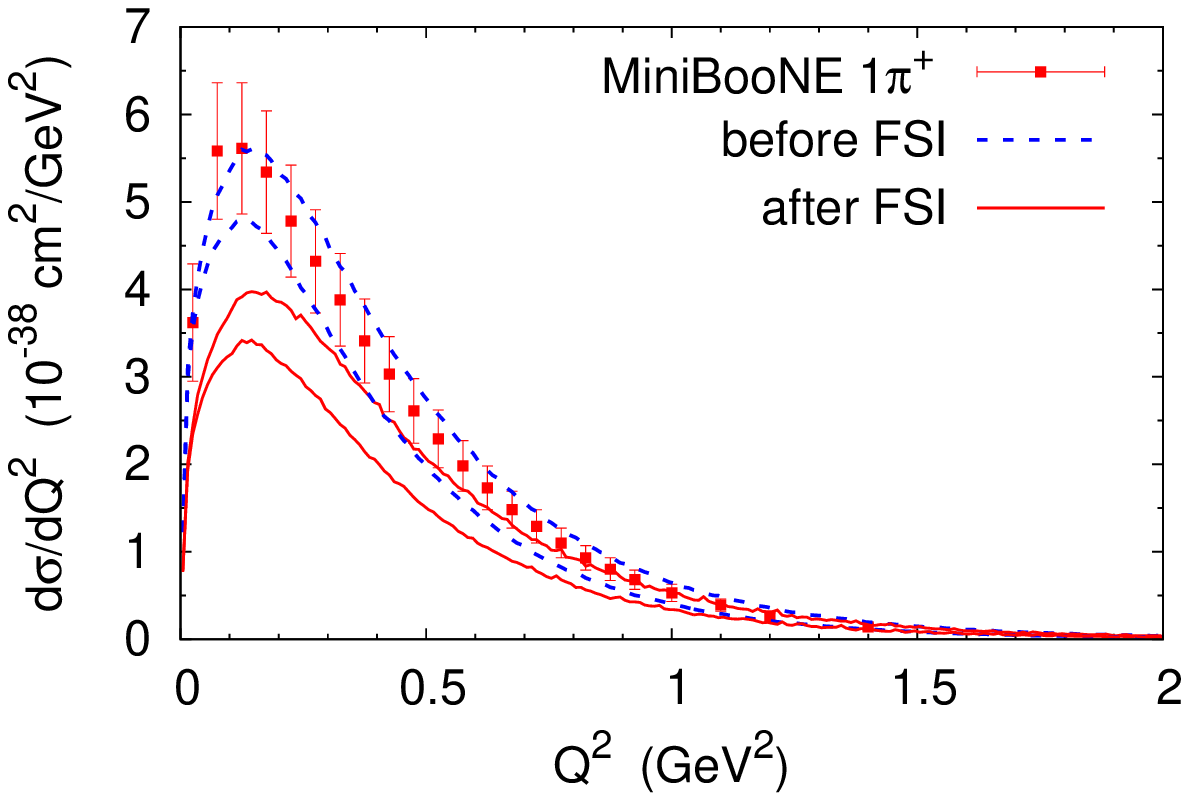}
\end{minipage}
\hfill
\begin{minipage}[c]{0.48\textwidth}
\includegraphics[width=\textwidth]{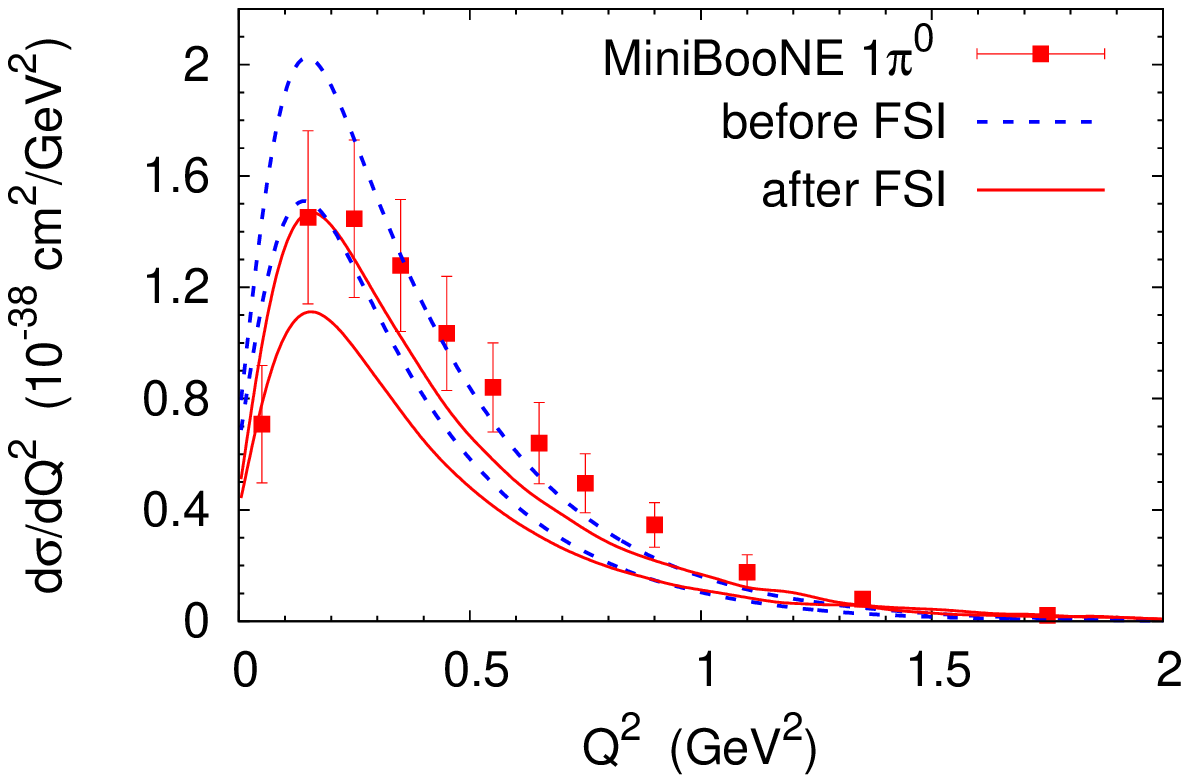}
\end{minipage}
\caption{(Color online) $Q^2$ distribution for the  1$\pi^+$ and 1$\pi^0$ production in the MiniBooNE.
Data are from~\cite{AguilarArevalo:2010bm,AguilarArevalo:2010xt}. The curves are as in Fig.\ \ref{fig:MB-lepton-Ekin}.}
\label{fig:MB-lepton-Q2}
\end{figure*}

In Figs.~\ref{fig:MB-lepton-Ekin}, \ref{fig:MB-lepton-coss} and \ref{fig:MB-lepton-Q2}
we compare these data with our calculations averaged over the published MiniBooNE flux.
As in the previous subsection, each cross section is calculated with ANL- and BNL-tuned elementary inputs, the corresponding curves
should be considered as lower and upper boundaries of the theoretical prediction.
For each distribution the results are given before (dashed lines)  and  after (solid lines) FSI.

As can be seen from these figures FSI hardly influence the shape of the distributions versus muon observables.
In particular, in the $\cos\theta_\mu$ distribution (Fig.~\ref{fig:MB-lepton-coss}) there is a slight suppression at forward angles and an enhancement at backward angles; correspondingly the $Q^2$ (Fig.~\ref{fig:MB-lepton-Q2})  distribution becomes somewhat flatter. The shape of the muon kinetic energy
distributions (Fig.~\ref{fig:MB-lepton-Ekin}) stays practically the same. This insensitivity is due to the fact that the only effect of FSI on these muon observables is that they can remove events in which an initially produced pion (or $\Delta$) was later on reabsorbed and bring in events in which the pion was produced only during FSI.

The curves after FSI are now compared with the experimental data.
For $1\pi^+$ production, when referring to the shape-only comparison, our curves correspond to the data. The muon kinetic energy distribution is consistently 20\% (upper boundary) to 60\% (lower boundary) lower than the data, see Fig.~\ref{fig:MB-lepton-Ekin}. For the $Q^2$ distribution in Fig.~\ref{fig:MB-lepton-Q2} the lower boundary is consistently 60\% lower than the data, while the upper boundary is around 20\% lower and a bit (within experimental errors) flatter than the data.

For $1\pi^0$ production, the shape of the curves evidently differs from the shape of the data. In some kinematical regions (low $T_\mu$, forward angles, moderate $Q^2$) our theoretical bands are below the data, while in other regions (moderate $T_\mu$, backward angles, low $Q^2$) they correspond to the data.

The lower bounds (obtained with the ANL input) after FSI and even before FSI are significantly lower than the experimental data for all three distributions for both $1\pi^+$ and $1\pi^0$ events.  For example, for $\pi^+$ the calculated $\dd \sigma/\dd T_\mu$ cross section after FSI amounts (at the peak) to only about 65\% of the measured value.
Even before the sizable pion-FSI the calculated cross section is below the data. For $\pi^0$ events, even before FSI the same cross section  amounts to only 65\% of the data; the one after FSI amounts to 55\%. In this case the FSI have  a relatively small effect. This is due to charge exchange that counteracts the pion absorption in this channel.  In line with this finding is the observation that FSI only slightly change the shape of the curves for $1\pi^0$ events.

\subsection{Pion observables}

\begin{figure*}[!hbt]
\begin{minipage}[c]{0.48\textwidth}
\includegraphics[width=\textwidth]{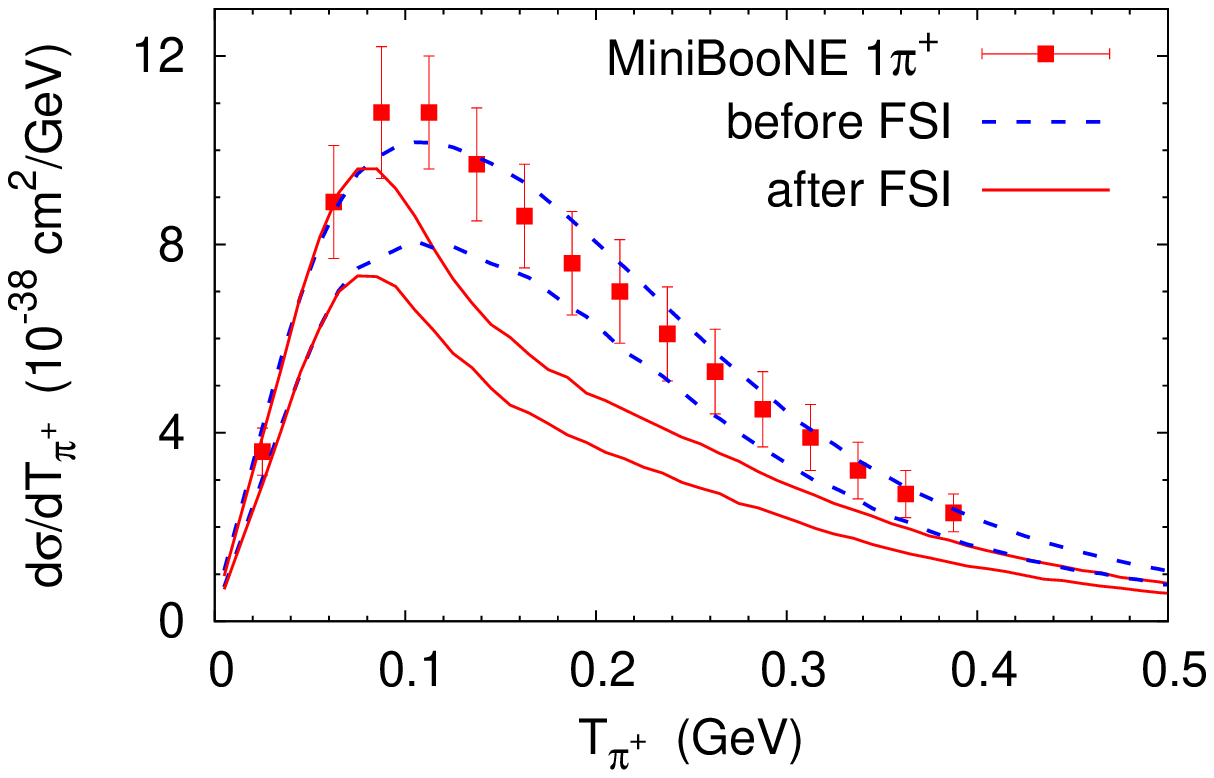}
\end{minipage}
\hfill
\begin{minipage}[c]{0.48\textwidth}
\includegraphics[width=\textwidth]{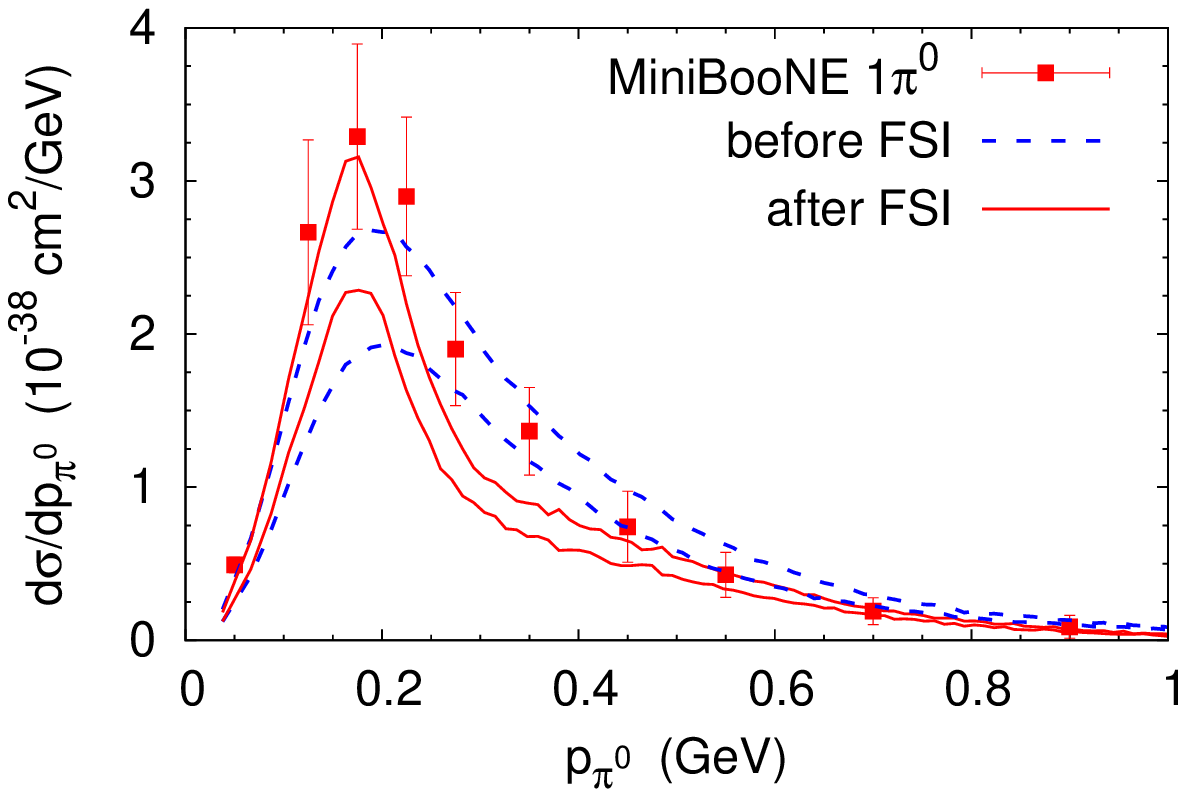}
\end{minipage}
\caption{(Color online) Kinetic energy distribution of the outgoing $\pi^+$ and momentum distribution of
the outgoing $\pi^0$ for \onepion production at MiniBooNE.
Data are from~\cite{AguilarArevalo:2010bm,AguilarArevalo:2010xt}. The curves are as in Fig.\ \ref{fig:MB-lepton-Ekin}.}
\label{fig:MB-pion-dTkin}
\end{figure*}

Figure~\ref{fig:MB-pion-dTkin} presents the results of our calculations for the kinetic energy distribution of the
outgoing pions. Similar to the muon-related distributions, the lower boundaries of our cross sections are significantly lower than the data. The upper boundaries again gives results closer to the data. However, for both charge states the shape of the calculated and measured distributions is very different; this disagreement we will discuss below.

\begin{figure*}[!hbt]
\begin{minipage}[c]{0.48\textwidth}
\includegraphics[width=\textwidth]{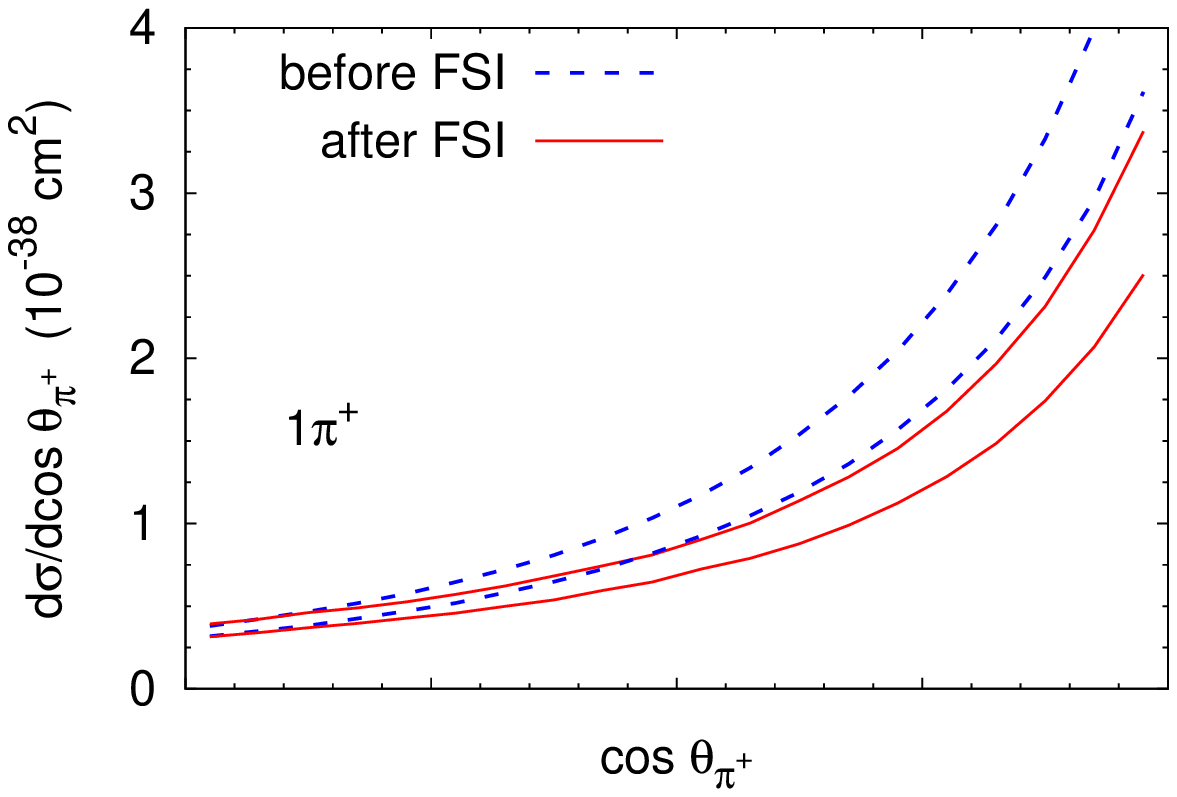}
\end{minipage}
\hfill
\begin{minipage}[c]{0.48\textwidth}
\includegraphics[width=\textwidth]{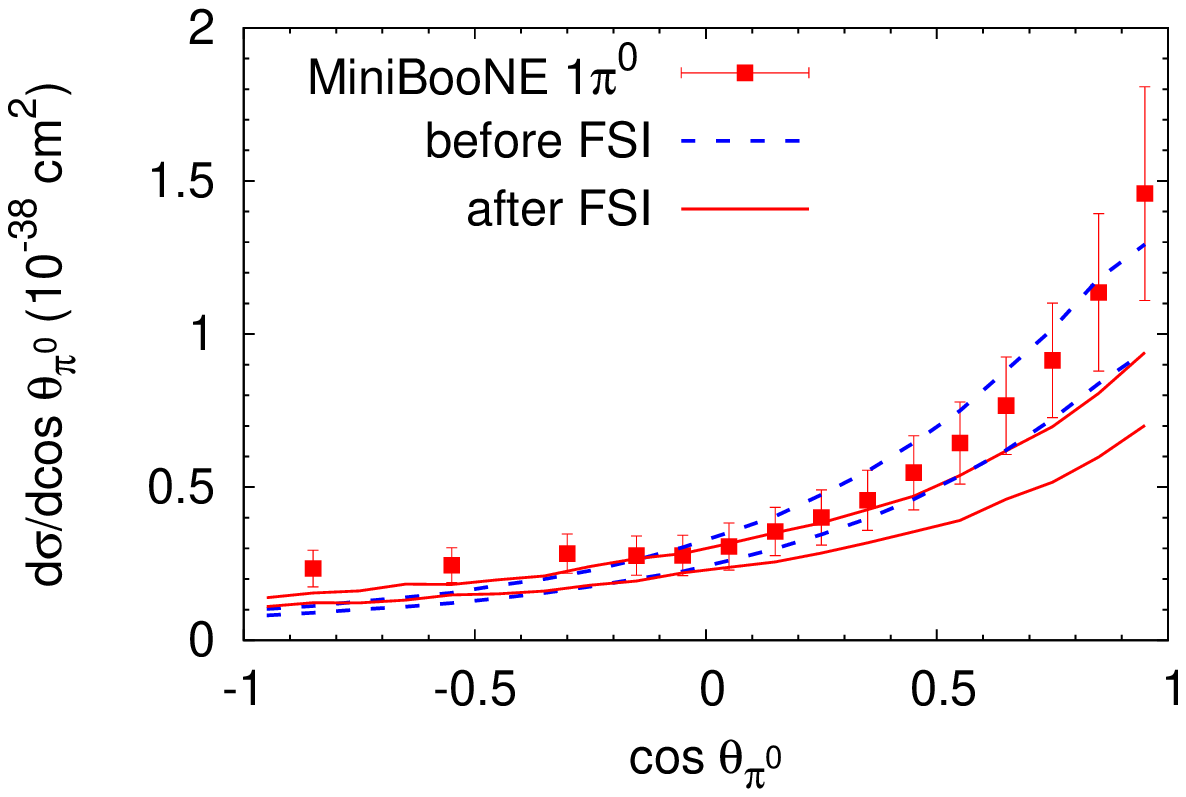}
\end{minipage}
\caption{(Color online) Distribution of the outgoing $\pi^+$  and $\pi^0$ in their angle relative to neutrino beam  for \onepion production at MiniBooNE.
Data are from~\cite{AguilarArevalo:2010xt}. The curves are as in Fig.\ \ref{fig:MB-lepton-Ekin}.}
\label{fig:MB-pion-dcospion}
\end{figure*}

Figure~\ref{fig:MB-pion-dcospion} shows the pion angular distribution of the CC $1\pi^+$ and $1\pi^0$
production.  Here data are available for $\pi^0$ only; for forward scattering
they are significantly higher than the lower boundary of our calculations, and only slightly higher compared to the upper boundary. The forward peaking is reproduced, but it is not as strong as exhibited by the data.

Unlike the case of muon-related observables, for pion-related distributions the FSI noticeably change their shape.
Surprisingly, the shape before FSI (see Fig.\ \ref{fig:MB-pion-dTkin}) is similar to that observed in the data. However, the shape of the calculated distributions after FSI looks markedly different. In particular, there is a significant lowering around $T_\pi \approx$ 0.2 GeV for $\pi^+$ and at around $p_{\pi}^0$ around 0.3 GeV, as a direct consequence of the $\Delta\pi N$ dynamics in nuclei. The following processes are important for this structure: pion elastic scattering in the FSI decreases the pion energy, thus depleting spectra at higher energies and accumulating strength at lower energies. Simultaneously, there is charge exchange scattering. At the same time pions are mainly absorbed via the $\Delta$ resonance, that is through $\pi N \to \Delta$ followed by $\Delta N \to N N$, which leads to the reduction in the region of pion kinetic energy $0.1-0.3 \GeV$. For $\pi^0$ production the additional increase of the cross section at lower energies comes from the side feeding of the $\pi^0$ channel from the dominant $\pi^+$ channel due to the charge exchange scattering $\pi^+  n \to \pi^0 p$. Inverse feeding $\pi^0 p \to \pi^+ n$ is suppressed, because at the energies under consideration, about 5 times less $\pi^0$s than $\pi^+$s are produced. The change of the shape of the spectra due to FSI is similar to that calculated for neutral current 1$\pi^0$ production in \cite{Leitner:2008wx}.

The particular shape calculated here for the neutrino-induced pions is in line with that observed experimentally in $(\gamma, \pi^0)$ production on nuclear targets \cite{Krusche:2004uw} (cf.\ Fig.\ \ref{fig:Lehr-pi0}). Since the shape depends on FSI and since the FSI are the same in both neutrino-induced and photon-induced reactions the absence of this special shape in the neutrino data is surprising.

\begin{figure*}[!hbt]
\begin{minipage}[c]{0.48\textwidth}
\includegraphics[width=\textwidth]{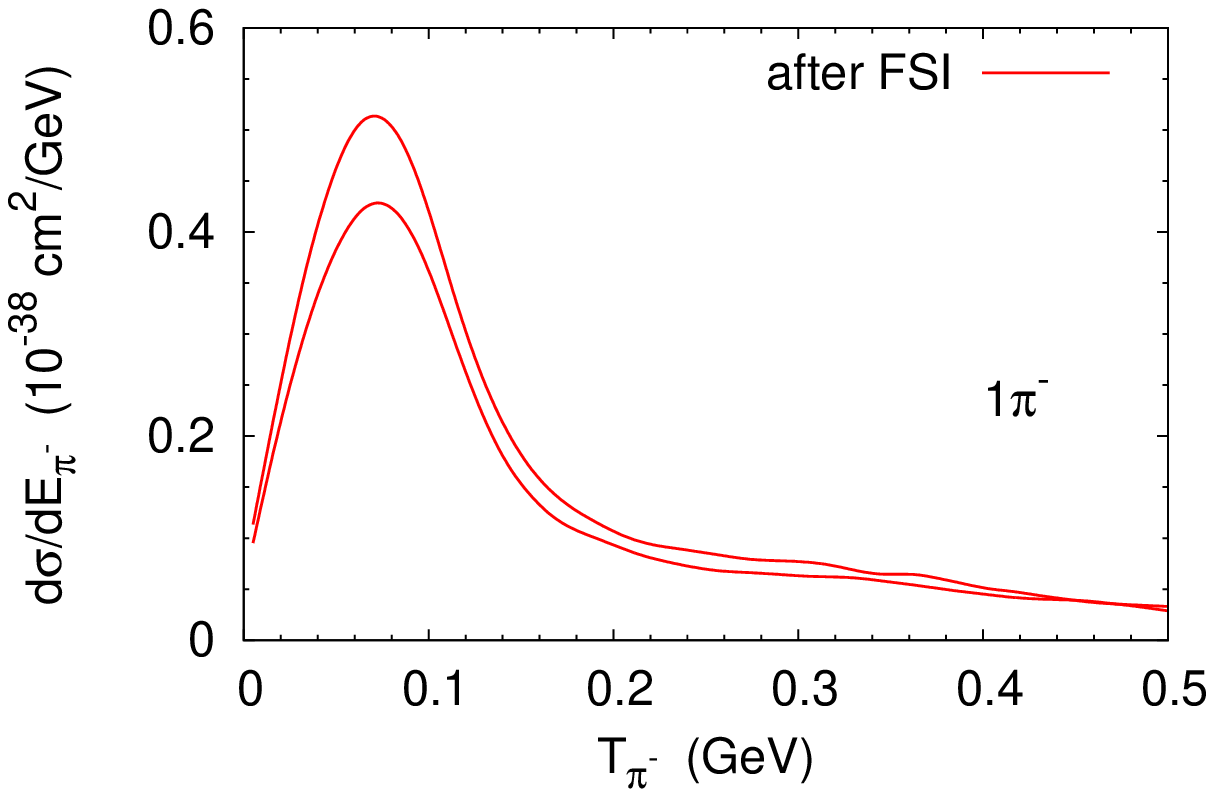}
\end{minipage}
\hfill
\begin{minipage}[c]{0.48\textwidth}
\includegraphics[width=\textwidth]{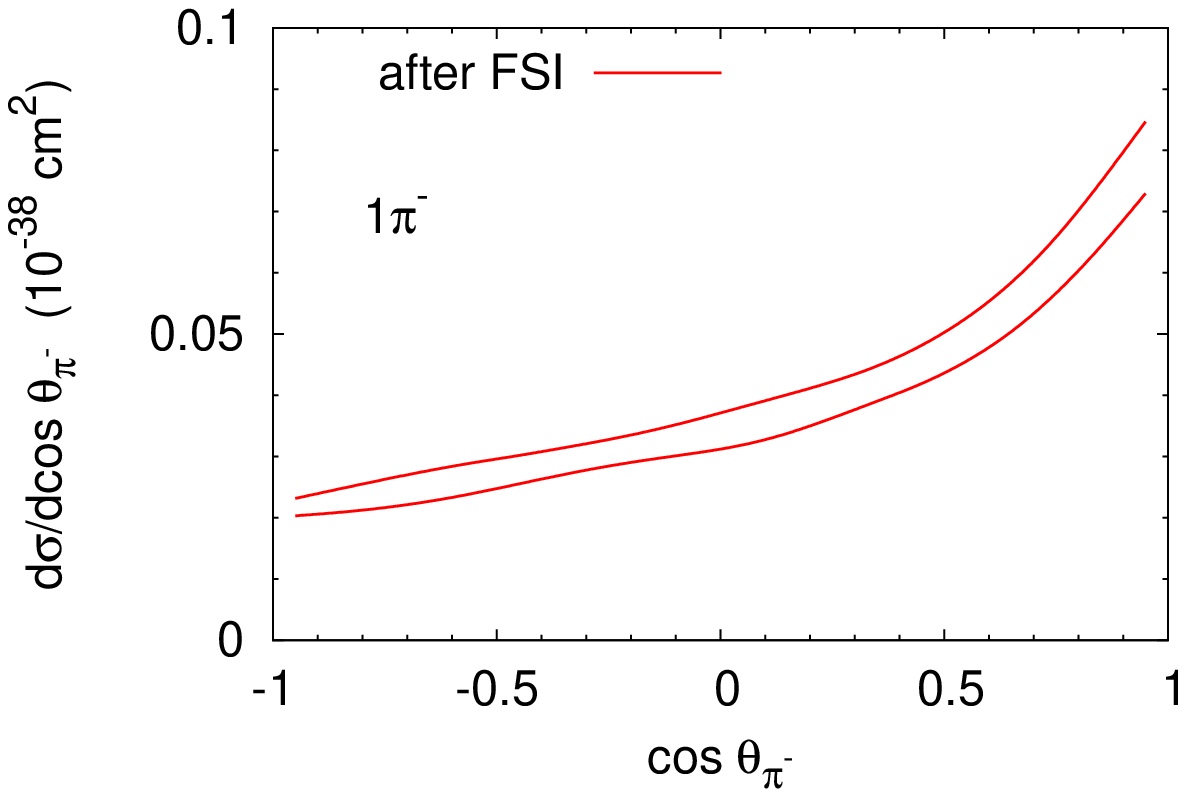}
\end{minipage}
\caption{(Color online) Distribution of the outgoing $\pi^-$ versus their kinetic energy and
 their angle relative to the neutrino beam for the MiniBooNE. The upper curves gives the result with the BNL input, the lower one that with the ANL input.}
\label{fig:MB-piminus}
\end{figure*}
The channel $\pi^0 n \to \pi^- p$ is important for $\pi^-$ production in the FSI.
Distributions for $\pi^-$ events are predictions of the GiBUU calculations; for completeness
they are shown in Fig.~\ref{fig:MB-piminus}. It is
noticeable that the angular distributions for $\pi^-$ are considerably less
forward-peaked than for the other pion charge states. This reflects the fact
that nearly all of the $\pi^-$ mesons are created by FSI.

%%%%%%%%%%%%%%%%%%%%%%%%%%%%%%%%%%%%%%%%%%%%%%%%%%%%%%%%%%%%%%%%%%%%%%%%%%%%%%%%%%
%%%%%%%%%%%%%%%%%%%%%%%%%%%%%%%%%%%%%%%%%%%%%%%%%%%%%%%%%%%%%%%%%%%%%%%%%%%%%%%%%%%

\section{Discussion of Results}
\label{sect:discreps}
Here we now summarize the comparison with experiment (cf.\ Fig. \ref{fig:MB-lepton-Enu-QEDelta}). For the ANL input, already the results for the energy unfolded \emph{total} $1\pi^+$ production cross sections \emph{without} FSI lie clearly (by $\approx 20$\%) \emph{below} the data;
those with FSI included are about 40-60\% below the experimental data. For the BNL input, the latter discrepancy still amounts to about 15-20\% for all energies, close to experimental uncertainties.

For 1$\pi^0$ production one observes a similar discrepancy: the lower boundary (ANL input) after FSI is considerably lower than the data; only for the highest energies the calculated curve lies within the error bars. The upper boundary (BNL input) is lower than the data for neutrino energies below $1.15\GeV$. At higher energies the theoretical curve is within experimental uncertainties and even rises with energy a bit steeper than the data.

It is interesting to note that the neutrino generator NUANCE, before any tuning, also gives cross sections that are consistently below the data. This can be seen, e.g.\ in Figs.\ 20 - 23 in \cite{AguilarArevalo:2010bm} and Figs.\ 8 - 14 in \cite{AguilarArevalo:2010xt}. Here for $\pi^+$ the calculations account only for about 80\% of the measured values, whereas for $\pi^0$ the discrepancy is much larger; for this channel only about 60\% of the experimental value is obtained. This is qualitatively consistent with the results presented here.

For the differential flux averaged cross sections  as a function of $T_\pi$ and of $\theta_\pi$ a more detailed picture emerges. Here it is seen that for both charged and neutral pions for kinetic energies up to about 60 MeV the data are very well reproduced.
However, above this kinetic energy the calculated cross sections are significantly lower than the data.
As discussed before the lowering of the cross section in this region is due to absorptive processes through the $\Delta$ resonance. It has been observed in photonuclear reactions and its absence in the neutrino-induced pion production is, therefore, unexpected.

\begin{figure*}[!hbt]
\begin{minipage}[c]{0.48\textwidth}
\includegraphics[width=\textwidth]{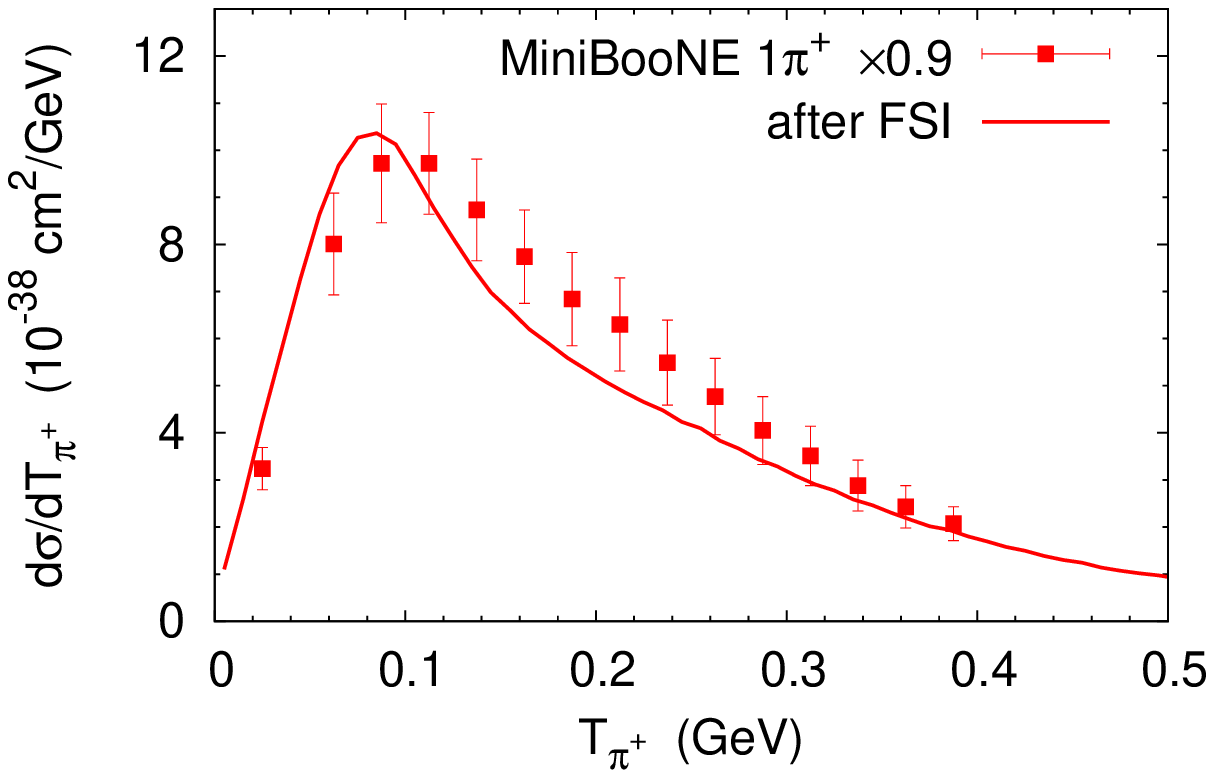}
\end{minipage}
\hfill
\begin{minipage}[c]{0.48\textwidth}
\includegraphics[width=\textwidth]{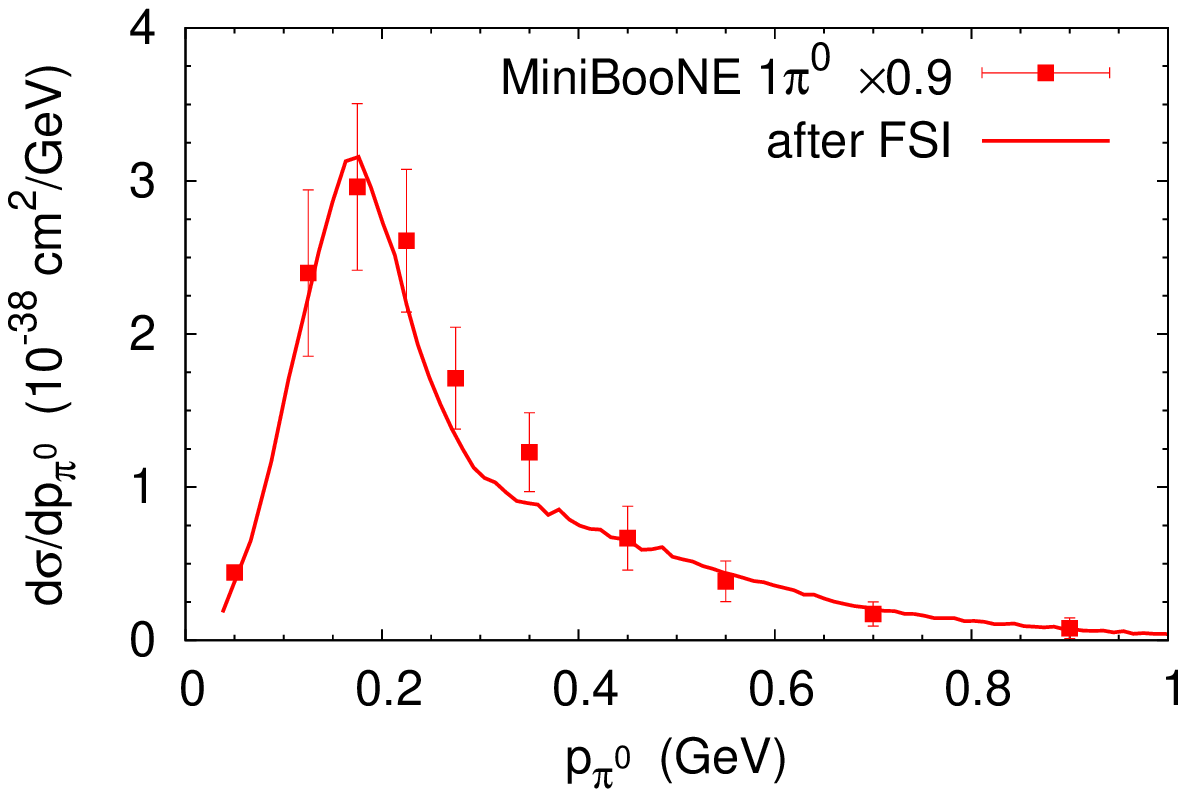}
\end{minipage}
\caption{(Color online) Left panel: Kinetic energy distribution of  the outgoing $\pi^+$. Right panel: momentum distribution of
the outgoing $\pi^0$. Data are from~\cite{AguilarArevalo:2010bm,AguilarArevalo:2010xt}, but multiplied with a factor 0.9.}
\label{fig:MB-pi+pi0fluxup}
\end{figure*}
An uncertainty contributing to the remaining discrepancy could
be a too low neutrino flux assumed by MiniBooNE in obtaining these pion data.
In \cite{AguilarArevalo:2008yp} the flux and energy spectrum have
been carefully determined. Different from other experiments the flux
has been obtained from a \textsc{GEANT} simulation of the primary hadronic
beam production cross sections for pions and kaons and their decay
into neutrinos. The remaining uncertainties are estimated to be of the order of 10 \%.
In the analysis of the quasielastic data there is indeed some indication that the flux assumed by MiniBooNE
may be too low by this amount \cite{Nieves:2011yp}. A flux correction by this amount would bring the upper boundary of our calculations (BNL-tuned input)
into better agreement with the data for $\pi^+$, except for the region around $T_\pi \approx 0.2$ GeV, where the special shape discussed above is not present in the data. For $\pi^0$ the calculations are with this flux renormalization within the experimental errors (see Fig.\ \ref{fig:MB-pi+pi0fluxup}).

We note that other theoretical calculations and in particular comparisons with the MiniBooNE data are extremely sparse. We have already pointed out that the results obtained with the NUANCE generator are qualitatively similar to our results obtained with the ANL input: a large underestimate of the measured $\pi^0$ yield and a much smaller underestimate for $\pi^+$. We are not aware of any other comparisons of CC pion spectra in the literature. For NC there exists a comparison \cite{Zeller2011} with results obtained both with a tuned version of NEUT and with a result obtained within NuWro \cite{Golan:2012wx} both of which describe the data. While the 'inner workings' of the former are not known to us, the latter model uses an unconventional implementation of formation times for pion production through the $\Delta$ resonance and thus leads to an underestimate of pion absorption. Finally we mention the calculations of Martini \emph{et al.}\ \cite{Martini:2009uj}. These authors calculated in an advanced model that included RPA correlations and many-particle interactions also total pion cross sections. While the calculated values seem to be of the right magnitude a detailed comparison is not possible since these calculations contain no final state interactions. The latter also is true for a recent approach extending the superscaling approximation to pion production \cite{Ivanov:2012fm}.

Summarizing this section we conclude that the $1\pi^+$ data obtained by MiniBooNE seem to suggest that the higher elementary BNL data for pion production are correct and that the ANL data underestimate the elementary production cross section. After a possible flux renormalization by about 10\% the calculations based on the BNL data are compatible with the MiniBooNE data on pion production. A discrepancy with theory still exists regarding the pion momentum distributions. Here the calculations, in agreement with experimental results on photoproduction of pions on nuclei, predict a suppression in the pion spectra at about the $\Delta$ resonance which is not seen in the published neutrino data.

\subsection{Many-body mechanisms in pion production}
All the possible explanations discussed so far still use the impulse approximation in which the incoming neutrino only interacts with one nucleon at a time. Many-body interactions are only taken into account through the collisional broadening of the $\Delta$ resonance. The latter is well established in photo- and electro-nuclear reactions and has, therefore, also been included here. If the actual reaction mechanism is more complicated than this and involves explicit many-body interactions then a component is missing in our calculations. In the experiment, the energy reconstruction would be affected and lead to wrong results. A similar situation has indeed been found to exist for quasielastic scattering where events identified experimentally as quasielastic scattering instead contained a significant contribution of 2p-2h events
\cite{Martini:2010ex,Martini:2009uj,Amaro:2011qb,Nieves:2011pp,Lalakulich:2012ac,Nieves:2012yz}. A corresponding mechanism leading to pion production would be, e.g.,  the process $\nu + N + N \to \mu + N + \Delta$ with the $\Delta$ finally decaying into $N \pi$. Such processes are not explicitly included in our calculations. The 2p-2h contributions to QE scattering become relevant in the dip region and under the $\Delta$ resonance.  It is, therefore, natural to expect a similar process, with one outgoing nucleon replaced by an outgoing $\Delta$ to become relevant at energies that are about 300 MeV higher than those where 2p-2h becomes relevant for QE. Room for such a process could thus be in the region of larger energy transfers beyond the $\Delta$ excitation. It is, however, a theoretical challenge to cleanly separate these processes from successive scattering in a transport description in order to avoid double counting.

Another process that could lead to larger pion cross sections is the production of off-shell pions. From detailed studies of the $\pi-N-\Delta$ dynamics it is well known that the pion potential in nuclear matter is for larger momenta attractive. Thus, one could expect an increase in initial (before FSI) pion production. However, first, the lowering of the pion mass goes together with an increase of the $\Delta$ mass; thus any pion production through the resonance would be suppressed. Second, the off-shell pion has to be propagated out of the nucleus. Studies of the effect of pion selfenergies on the pion production cross sections in heavy-ion reactions have given the result that the final, observable pion yield is hardly affected by the presence of a pion potential \cite{Ehehalt:1993px,Xiong:1993pd,Larionov:2002jw} because in dense matter most pions are absorbed and the last interaction of the pion happens for rather low densities $\rho < 0.5 \rho_0$ where the selfenergies are small. For neutrino-induced reactions on nuclei this point remains to be investigated.

An estimate of an upper limit for the importance of explicit many-body effects could be obtained from a closer inspection of the pion photoproduction data in
Fig.\ \ref{fig:Lehr-pi0} where the theoretical prediction has been obtained without any explicit many-body production mechanisms; only the collisional width of the $\Delta$ resonance has been used. This figure shows that the total cross section obtained is about 20\% too low at photon energies around 500 MeV. This number thus sets a limit to any possible many-particle production processes, assuming that their importance is similar for vector and axial couplings.

\section{Conclusion}
\label{conclusion}

In this paper we have calculated the neutrino-induced pion production on a nuclear target and have compared the results with the recent MiniBooNE data. The calculations are based on the impulse approximation in which the incoming neutrino interacts with only one nucleon at a time. We have found that the experimentally measured total cross sections are always higher than the theoretical ones. However, there exists a strong dependence of the calculated result on the elementary cross sections used as input. The latter are not well constrained; the higher data set (from the BNL experiment) gives total theoretical cross sections closer (within $\approx 20 \%$) to the data.

An inspection of the pion spectra shows that the difference between theory and experiment can be attributed to the pion momentum distributions which in the theoretical calculations show a clear effect of pion absorption through the $\Delta$ resonance; this effect is missing in the data.

A final comparison, in which the experimental flux has been renormalized by 10\%, gives results for the differential cross sections that are generally in quite good agreement with the MiniBooNE data, when the BNL elementary data are used as input. The discrepancy in the pion spectra at a pion momentum of about 0.3 GeV is still there and more pronounced for $\pi^+$ production. However, in both cases the calculated results lie close to the lower end of the experimental error bars so that the disagreement may not be statistically well established. This result relies on the use of the in-medium collision-broadened width of the $\Delta$; without it the cross sections would be significantly larger in the peak region and the spectral shape would be worse.

This leaves us with the unsatisfactory conclusion that there are two possibilities to explain the data.

\begin{itemize}
\item First, the BNL data describe the elementary cross sections correctly. Then, together with a flux renormalization within the experimental uncertainties, the data obtained by MiniBooNE can mostly be explained without invoking any many-body production processes. Without flux renormalization many-body effects could amount to about 10 - 20\% at most.
\item Second, the ANL data describe the elementary cross sections correctly. In this case the nuclear production cross sections are clearly too low and sizable many-body production effects would have to be invoked to bring the calculated cross section up to the measured values. A change of the flux within its experimental uncertainty would not be sufficient. Instead, many-body effects would be needed to generate about 40\% of the cross section.
\end{itemize}

Even though the first alternative sounds more convincing to us, in the present situation it is not possible to make a definite choice between these possibilities. Only a remeasurement of the elementary cross sections would clarify the situation. With reliable elementary cross sections at hand one would have a clear handle on the importance of many-body effects in neutrino-induced pion production on nuclei.

\acknowledgments
This work is supported by DFG and BMBF. We gratefully acknowledge many helpful comments on the pion data by R. Nelson and G. Zeller. U.M. is grateful for many helpful discussions on neutrino-induced pion production with J. Nieves and L. Alvarez-Ruso. We also have benefited from comments by J. Sobzcyk clarifying for us the details of pion production in the generator NuWro.

\appendix

\section{Pion production through the $\Delta$ resonance}
\label{app:pion-theory}
For easier reference we collect here briefly the essentials of our calculation of the pion production cross section $\dd \sigma_{1p1h1\,\pi}^{\rm med}$
in Eq.\ (\ref{IA}). For further details we refer to Ref.\ \cite{Leitner:2008ue}.

The cross section for the resonance excitation in the reaction $\ell(k) N(p) \to \ell'(k') R(p')$ is in general given
\begin{equation}   \label{X-sec}
\frac{\dd \sigma_R}{\dd \omega \dd\Omega'} = \frac{|\mathbf{k}'|}{32 \pi^2} \frac{\mathcal{A}(p'^2)}{\left[(k \cdot p)^2 - m_l^2M^2\right]^{1/2}} |\bar{\mathcal{M}}_R|^2 ~.
\end{equation}
Here the spectral function $\mathcal{A}$ is given by
\begin{equation}    \label{Deltaspectr}
\mathcal{A}(p'^2) = \frac{\sqrt{p'^2}}{\pi} \frac{\Gamma(p')}{(p'^2 - M_R^2)^2 + p'^2\Gamma^2(p')} ~.
\end{equation}
The free resonance width $\Gamma$ is momentum dependent; it is obtained from the Manley analysis \cite{Manley:1992yb} with a pole value of 0.118 GeV.

The matrixelement in (\ref{X-sec}) is given by
\begin{equation}
|\bar{\mathcal{M}}_R|^2 = G_F \frac{\cos\theta_C}{\sqrt{2}} L_{\mu \nu} H_R^{\mu \nu}~,
\end{equation}
where $\theta_C$ is the Cabibbo angle. The hadronic tensor for resonance excitation on one nucleon, $H_R$, is
\begin{equation}
H_R^{\mu \nu} = \frac{1}{2} {\rm Tr} \left[(\slashed{p} + M)\, \tilde\Gamma^{\alpha \mu} \Lambda_{\alpha \beta}\Gamma^{\beta \nu}\right] ~.
\end{equation}
Here $\Lambda$ is, for the $\Delta$ resonance, the spin 3/2 projector.  Vertex $\Gamma$ of weak nucleon-resonance interactions involves phenomenological form factors, $\tilde \Gamma=\gamma^0 \Gamma^\dagger \gamma^0$. Its vector part is completely determined by electron scattering. Its axial part, on the other hand, contains the crucial form factor generally denoted by $C_5^A$ (for more details see sec.\ II.B.2 in \cite{Leitner:2008ue}). The resonance contribution to the pion production cross section $\dd\sigma_{1p1h\,1\pi}$ is then obtained by multiplying $\dd \sigma_R$ with the free branching ratio for decay into $\pi N$ and assuming an isotropic angular distribution in the $\Delta$ restframe.

The form factor $C_5^A$ has to be determined by a fit to elementary pion production data. The ANL data for $\dd\sigma /\dd Q^2$ \cite{Barish:1978pj,Radecky:1981fn} have been fitted in \cite{Leitner:2008ue} by
\begin{equation}
C_5^A(Q^2) = C_5^A(0) \left[1 + \frac{aQ^2}{b + Q^2}\right] \left(1 + \frac{Q^2}{{M_A^\Delta}^2}\right)^{-2}
\end{equation}
with a = - 1.21 and b = 2 GeV$^2$, with an axial mass of $M_A^\Delta\mbox{(ANL)} = 0.95$ GeV and $C_5^A(0) = 1.17$, consistent with PCAC.

We have now fitted also the BNL data with the same functional form. The fit parameters differ from those for the ANL data only by the axial mass which is larger: $M^\Delta_A\mbox{(BNL)} = 1.3$ GeV.

While these fits determine the resonance contribution the necessary background terms have been determined following again \cite{Leitner:2008ue}. There the vector part $\sigma^V_{\rm BG}$ was again determined by a fit to electron data, using the MAID analysis as a basis. For the axial part (including the vector-axial interference) it was then assumed, for simplicity, that it is proportional to the vector part, so that the total cross section then is given by
\begin{equation}
\dd \sigma_{\rm BG} = (1 + b^{N\pi})\, \dd \sigma^V_{\rm BG} ~,
\end{equation}
The total pion production cross section for the $1p1h\,1\pi$ final state is then given by a sum of resonance and background contribution where the latter also includes the resonance-background interference term
\begin{equation}
\sigma_{1p1h\,1\pi} = \sigma_R\frac{\Gamma_\pi}{\Gamma_{\rm tot}} + \sigma_{\rm BG} ~.
\end{equation}
A fit to the ANL data was achieved with $b^{p\pi^0}\mbox{(ANL)} = 3$ and $b^{n\pi^+}\mbox{(ANL)} = 1.5$. The fit to the BNL data has given the values $b^{p\pi^0}\mbox{(BNL)} = 6.0$ and $b^{n\pi^+}\mbox{(BNL)} = 3.0$.

All calculations in the present paper have been done with these parameters. The curves in Fig.\ \ref{fig:ANLBNL1pidata} contain not only the resonance and background contributions described here, but also contributions from higher resonances and DIS; the latter start to contribute for neutrino energies above about 1 GeV.

\section[Influence of medium modification of the $\Delta$ ]{Influence of medium modification of the Delta}
\label{app:Oset}

In all of the calculations in this paper we have used a medium modification of the $\Delta$  width according to the Oset and Salcedo (OS) model~\cite{Oset:1987re}.
In medium the width $\Gamma$ in Eq.\ (\ref{Deltaspectr}) acquires a collisional contribution which is given by
\begin{equation}
\Gamma_{\rm coll} = - \frac{2}{\sqrt{p'^2}} \mathrm{Im} \Sigma_{\rm coll}(p'^2)  ~,
\end{equation}
where the selfenergy $\Sigma_{\rm coll}$ is a function of density and momentum (see Eq.\ (4.4) in \cite{Oset:1987re}). In the neutrino reactions investigated here the momenta of the $\Delta$ resonances produced can range up to $\approx 1.5$ GeV, i.e.,\ well above the range of validity of the OS model where $p_\Delta < 0.3$ GeV. As a consequence, use of this $\Delta$ width requires a significant amount of extrapolation for the higher momenta. For simplicity, we freeze the collisional width at the highest calculated value. The price one has to pay for this is a loss of spectral strength. At the higher momenta about 10 - 15\% of the spectral strength is missing (see Table 7.5 and discussion in \cite{Leitner:2009zz}). All calculations in this paper, except otherwise noted, have been done using this collisional width.

Here we illustrate in some more detail how this collisional broadening influences the cross sections. Since the OS model involves many-body effects on the $\Delta$ collisional width it is essential that the parametrization used is consistent with the actual collisions in GiBUU. This is indeed the case; in particular, when using the OS model then also 3-body collision terms are automatically turned on.

\begin{figure}[bht]
\includegraphics[width=\columnwidth]{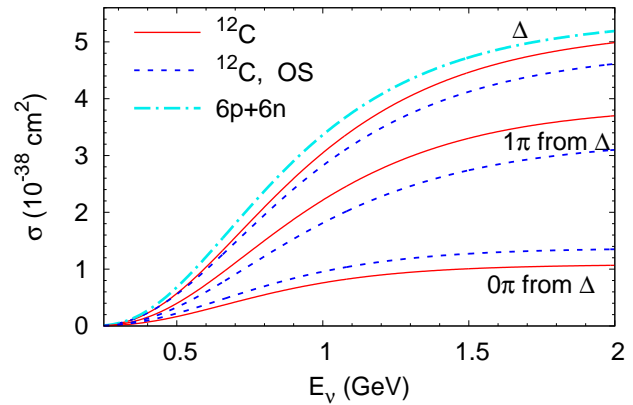}
\caption{(Color online) Cross sections for $\Delta$ production, $1\pi$-production originating from $\Delta$
and  $0\pi$-production originating from $\Delta$  with and without the OS medium modification \cite{Oset:1987re}.}
\label{fig:carbon-Delta-1pi-nopi}
\end{figure}

The influence of initial state interactions and medium modifications of the $\Delta$ resonance on the $\Delta$ production
cross section for neutrino scattering off carbon nucleus is illustrated in Fig.~\ref{fig:carbon-Delta-1pi-nopi}.
The cross section for 6 free neutrons and 6 free protons, all at rest,
is shown as dash-dotted curve labeled ``6p+6n''. It provides a reference for the following discussions. The nuclear effects,
namely the Fermi motion, the binding nuclear potential and the Pauli blocking of the outgoing proton,
decrease the cross section by around $5\%$, which is shown as solid curve labeled ``${}^{12}$C''.
Taking into account the OS modification of Delta properties leads to an additional decrease of the cross section by another $5-8\%$.
This is shown as dashed curve labeled ``${}^{12}$C, OS''.
This decrease is a consequence of the shift of strength from the $\Delta$ peak position to larger masses due to the increased broadening. This extra strength at the higher masses is, however, cutoff by the form factors so that the net effect is a lowering of the cross section.

Since the free $\Delta$ primarily decays to a pion and a nucleon, one would expect a similar decrease for the \onepion cross section.
As one can easily see, this is indeed the case, but the effect is larger: the cross section is decreased by $15-20\%$. This additional suppression is due to the fact that the $\Delta$ now has an increased collision width so that it undergo $\Delta + N \to N N$ before the pion decay has taken place.

\begin{figure}[bht]
\includegraphics[width=\columnwidth]{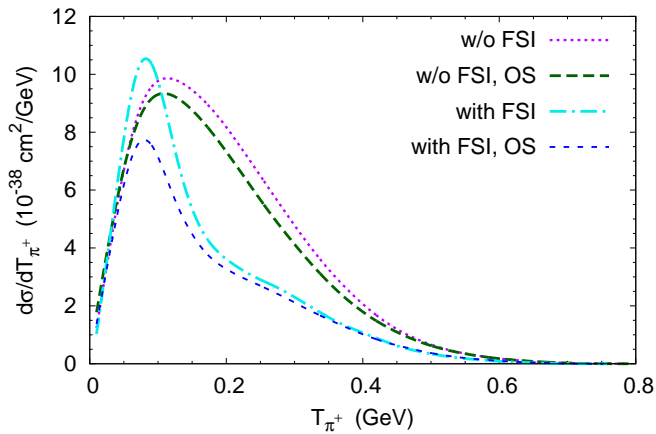}
\caption{(Color online)  Kinetic energy distribution of $\pi^+$ produced in neutrino scattering off carbon
 through the weak production of the $\Delta$ resonance and its following decay. The neutrino energy is $E_\nu=1\GeV$. The curves labeled OS were obtained using the in-medium collisional width of the $\Delta$ from \cite{Oset:1987re}.}
\label{fig:carbon-pionplus}
\end{figure}

While Fig.\ \ref{fig:carbon-Delta-1pi-nopi} shows the results for the initial $\Delta$ production, which does not depend on FSI, we now
consider the influence of FSI on the charged pions, which originates from the $\Delta$ resonance production.
Fig.~\ref{fig:carbon-pionplus} shows the kinetic energy energy distribution of these $\pi^+$.
Before FSI the distribution has a rather broad peak at $T\approx 0.1 \GeV$,
which is decreased by about $5\%$, if the OS modification of the $\Delta$ properties is taken into account.

After FSI, the OS modification of the $\Delta$ properties decreases the pion distribution significantly more,
by more than $20\%$ at the peak. This is due to the fact that together with the collisional broadening of the $\Delta$ resonance explicit two-body and three-body collision terms are now active. Thus, before having a chance to decay to a pion
a significant part of $\Delta$s is absorbed in the nucleus.
The overall effect is a decreased pion production cross section which increases the number of pionless events and
thus serves as a source of the fake QE-like events.

\bibliographystyle{apsrev4-1}
\bibliography{nuclear}

\end{document}